\newif\ifarxiv\arxivtrue

\arxivtrue     

\ifarxiv
\documentclass[12pt,a4paper]{book}
\pdfoutput=1
\usepackage{titling} 
\usepackage[labelfont=bf]{caption}
\usepackage{amsthm}
\newtheoremstyle{examplestyle}{1   mm}{1mm}{\slshape}{6pt}{\bfseries}{\newline}{10mm}{}
\theoremstyle{definition}
\newtheorem{example0}{Exercise}

\usepackage[nopar]{lipsum}
\usepackage[x11names]{xcolor}

\newenvironment{exercise}{%
	\par\noindent{\color{DarkSeaGreen3}\hrule height 1pt}\nopagebreak\hfill\nopagebreak\vspace{1.3mm} \vspace*{-\baselineskip}\nopagebreak
	\begin{example0}%
	}{%
		\strut{\color{DarkSeaGreen3}\hrule height 1pt\hfill}
	\end{example0}
}

\makeatletter

\divide\@tempdima\baselineskip
\@tempcnta=\@tempdima
\makeatother

\else
\documentclass{pinchcrn}   
\fi

\usepackage[Lenny]{fncychap}
\usepackage{hyperref}
\usepackage{graphicx}
\usepackage[utf8]{inputenc}
\usepackage{amsmath}
\usepackage{framed}
\usepackage{color}
\definecolor{shadecolor}{rgb}{0.96,0.96,0.91}
\definecolor{shadecolor2}{rgb}{1,1,0.99}

\DeclareFixedFont{\ttb}{T1}{txtt}{bx}{n}{12} 
\DeclareFixedFont{\ttm}{T1}{txtt}{m}{n}{12}  

\newcommand{\la}[1]{\label{#1}}
\newcommand{\eq}[1]{(\ref{#1})}
\newcommand{\nn}{\nonumber}
\newcommand{\bP}{{\bf P}}
\newcommand{\bQ}{{\bf Q}}
\newcommand{\beq}{\begin{equation}}
\newcommand{\eeq}{\end{equation}}
\newcommand\beqa{\begin{eqnarray}}
\newcommand\eeqa{\end{eqnarray}} 
\newcommand\bea{\begin{array}}
\newcommand\eea{\end{array}}

\def\thebibliography#1{%
	\chapter*{\bibname}
	\addcontentsline{toc}{chapter}{\bibname}
	\chaptermark{\bibname}
	\bgroup\list{\arabic{enumi}.}
	{\settowidth\labelwidth{#1}
		\leftmargin\labelwidth
		\itemindent -\leftmargin
		\itemsep 0pt
		\parsep \itemsep
		\advance\leftmargin\labelsep 
		\usecounter{enumi}}}

\definecolor{deepblue}{rgb}{0,0,0.6}
\definecolor{deepred}{rgb}{0.6,0,0}
\definecolor{deepgreen}{rgb}{0,0.5,0}

\usepackage{listings}

\newcommand\pythonstyle{\lstset{
language=Python,
basicstyle=\ttm,
otherkeywords={self},             
keywordstyle=\ttb\color{deepblue},
emph={MyClass,__init__},          
emphstyle=\ttb\color{deepred},    
stringstyle=\color{deepgreen},
frame=tb,                         
showstringspaces=false            %
}}

\lstnewenvironment{python}[1][]
{
\pythonstyle
\lstset{backgroundcolor=\color{shadecolor}}
}
{}

\lstnewenvironment{pythonout}[1][]
{
\pythonstyle
\lstset{backgroundcolor=\color{shadecolor2}}
}
{}

\lstnewenvironment{pythonout2}[1][]
{\begin{lrbox}{\thmbox}%
\begin{minipage}{\dimexpr\linewidth-2\fboxsep}
}%
{\end{minipage}%
\end{lrbox}%
\begin{trivlist}
\item[]\colorbox{lightgray}{\usebox\thmbox}
\end{trivlist}}

\newcommand\pythoninline[1]{{\pythonstyle\lstinline!#1!}}
\newcommand\pyi[1]{{\pythonstyle\lstinline!#1!}}
\newcommand\pyin[1]{{\pythonstyle\lstinline!#1!}}
\newcommand\pline[1]{{\pythonstyle\lstinline!#1!}}

\newcommand\mathstyle{\lstset{
language=Mathematica,
basicstyle=\ttfamily\bfseries\small,
otherkeywords={self},             
keywordstyle=\ttfamily\bfseries\small\color{deepblue},
emph={MyClass,__init__},          
emphstyle=\ttfamily\bfseries\small\color{deepred},    
stringstyle=\color{deepgreen},
frame=tb,                         
showstringspaces=false            %
}}

\lstnewenvironment{mathematica}[1][]
{
\mathstyle
\lstset{#1}
}
{}

\newcommand\mline[1]{{\mathstyle\lstinline!#1!}}

\ifarxiv

\title{Introduction to the Spectrum of \texorpdfstring{${\mathcal{N}=4}$}{N=4} SYM and the Quantum Spectral Curve}
\author{Nikolay Gromov}
\date{\small Mathematics Department, King's College London,
	The Strand, London WC2R 2LS, UK. \& St.Petersburg INP, Gatchina, 188 300, St.Petersburg,
	Russia\\ \vspace{10mm}
	{\bf Abstract.}
{\it
This review is based on the lectures given by the author at the Les Houches Summer School 2016. It describes the recently developed Quantum Spectral Curve (QSC) for a non-perturbative planar spectrum of N=4 Super Yang-Mills theory in a pedagogical way starting from the harmonic oscillator and avoiding a long historical path. We give many examples and provide exercises. At the end we give a list of the recent and possible future applications of the QSC. }\\ \vspace{100mm}
{\bf Dedication.} In memory of Ludvig Dmitrievich Faddeev.
}

\begin{document}
\begin{titlingpage}
	\maketitle
\end{titlingpage}
\tableofcontents
\else

\title{Les Houches Lecture Notes:\\ Spectrum of N=4 SYM and the Quantum Spectral Curve}

\author{Nikolay Gromov}
\affiliation{Mathematics Department, King's College London,
The Strand, London WC2R 2LS, UK. \& St.Petersburg INP, Gatchina, 188 300, St.Petersburg,
Russia}

\begin{document}

\maketitle

\dedication{In memory of Ludvig Dmitrievich Faddeev.}

\acknowledgements

I am very grateful to M.Alfimov, A.Cavagli\`a, S.Leurent, F.Levkovich-Malyuk,   G.Sizov, D.Volin, and especially to V.Kazakov and P.Vieira for numerous discussions on closely related topics. I am thankful to D.Grabner, D.Lee and J.\footnote{i.e. Julius, who only has a first name} for carefully reading the manuscript.
The work was supported   by the European Research Council (Programme
``Ideas" ERC-2012-AdG 320769 AdS-CFT-solvable). We are grateful to Humboldt
University (Berlin) for the hospitality and financial support of this work in the framework of the ``Kosmos" programe. We wish to thank STFC for support from Consolidated grant number ST/J002798/1. This work has received funding from the People Programme (Marie Curie Actions)
of the European Union's Seventh Framework Programme FP7/2007-2019/ under REA Grant Agreement No 317089 (GATIS).\\
{ }\\
Please report typos or send other improvement requests for these lecture notes to \url{nikgromov@gmail.com}.

\tableofcontents

\maintext

\fi

\chapter{Introduction}
The importance of AdS/CFT correspondence in modern theoretical physics and the role of ${\mathcal N}=4$ SYM in it is hard to over-appreciate. 
In these lecture notes we try to give a pedagogical introduction to the Quantum Spectral Curve (QSC) of ${\mathcal N}=4$ SYM,
a beautiful mathematical structure which describes the non-perturbative spectrum of strings/anomalous dimensions of all single trace operators.
The historical development leading to the discovery of the QSC~\cite{Gromov:2011cx,Gromov:2014caa} is a very long and interesting story by itself, and there are several reviews trying to cover the main steps on this route \cite{Beisert:2010jr,Bombardelli:2016rwb}.
For the purposes of the lectures we took another approach and try to motivate the construction by emphasizing numerous analogies between the QSC construction and
 basic quantum integrable systems such as the harmonic oscillator, Heisenberg spin chains, and classical sigma-models. In this way the QSC comes out naturally, bypassing
extremely complicated and technical stages such as derivation of the S-matrix \cite{Beisert:2005tm}, dressing phase \cite{Janik:2006dc}, mirror theory \cite{Ambjorn:2005wa}, Y-system \cite{Gromov:2009tv}, Thermodynamic Bethe Ansatz 
\cite{Gromov:2009bc,Bombardelli:2009ns,Arutyunov:2009ur,Cavaglia:2010nm}, NLIE~\cite{Gromov:2011cx,Balog:2012zt} and finally derivation of the QSC~\cite{Gromov:2011cx,Gromov:2014caa}.

We also give examples of analytic solutions of the QSC and in the last chapter describe step-by-step the numerical algorithm allowing us to get the non-perturbative spectrum with almost unlimited precision~\cite{Gromov:2015wca}. 
We also briefly discuss the analytic continuation of the anomalous dimension to the Regge (BFKL) limit relevant for more realistic QCD.

The structure is the following: in the Chapter~1 we re-introduce the harmonic oscillator and the Heisenberg spin chains in a way suitable for generalization to the QSC.
Chapter~\ref{ch:2} describes classical integrability of strings in a curved background, which give some important hints about the construction of the  QSC. In Chapter~\ref{ch:3} we give a clear formulation of the QSC.
In Chapter~\ref{ch:4} we consider some analytic examples. And in the last Chapter~\ref{ch:5} we present the numerical method.
\newpage
\paragraph*{Acknowledgment}
I am very grateful to M.Alfimov, A.Cavagli\`a, S.Leurent, F.Levkovich-Malyuk,   G.Sizov, D.Volin, and especially to V.Kazakov and P.Vieira for numerous discussions on closely related topics. I am thankful to D.Grabner, D.Lee and J.\footnote{i.e. Julius, who only has a first name} for carefully reading the manuscript.
The work was supported   by the European Research Council (Programme
``Ideas" ERC-2012-AdG 320769 AdS-CFT-solvable). We are grateful to Humboldt
University (Berlin) for the hospitality and financial support of this work in the framework of the ``Kosmos" programe. We wish to thank STFC for support from Consolidated grant number ST/J002798/1. This work has received funding from the People Programme (Marie Curie Actions)
of the European Union's Seventh Framework Programme FP7/2007-2019/ under REA Grant Agreement No 317089 (GATIS).\\
{ }\\
Please report typos or send other improvement requests for these lecture notes to \url{nikgromov@gmail.com}.

\chapter{From Harmonic Oscillator to QQ-Relations}\la{ch:2}

\section{Inspiration from the Harmonic Oscillator}
To motivate the construction of the QSC we first consider the 1D harmonic oscillator and concentrate on
the features which, as we will see later, have similarities with the construction for the spectrum of ${\mathcal N}=4$ SYM.

The harmonic oscillator is the simplest integrable model which at the same time exhibits nontrivial features surprisingly similar to ${\mathcal N}=4$ SYM. Our starting point is the Schr\"odinger equation
\beq\la{SH}
-\frac{\hbar^2}{2m} \psi''(x)+V(x)\psi(x)=E\psi(x)
\eeq
where $V(x)=\frac{m\omega^2 x^2}{2}$.
Alternatively, it can be written in terms of the quasi-momentum
\beq\la{pdef}
p(x)=\frac{\hbar}{i}\frac{\psi'(x)}{\psi(x)}
\eeq
as
\beq\la{ppeq}
p^2-i\hbar p'=2m(E-V)\;.
\eeq
This non-linear equation is completely equivalent to \eq{SH}.
Instead of solving this equation directly let us make a simple ansatz for $p(x)$.
We see that for large $x$ the r.h.s. behaves as $-m^2\omega^2 x^2$
implying that at infinity $p\simeq i m \omega x$. Furthermore, $p(x)$ should have simple poles
at the position of zeros of the wave function which we denote $x_i$.
All the residues at these points should be equal to $\hbar/i$ as one can see from $\eq{pdef}$.
We can accommodate all these basic analytical properties with the following ansatz:
\beq\la{px}
p(x)=i m \omega x+\frac{\hbar}{i}\sum_{i=1}^N\frac{1}{x-x_i}\;.
\eeq
We note that at large $x$ the r.h.s. of \eq{px} behaves as $i m\omega x+\frac{\hbar}{i}\frac{N}{x}+{ O}(1/x^2)$.
Plugging this large $x$ approximation of $p(x)$ into the exact equation
\eq{ppeq} we get:
\beq
\left(i m\omega x+\frac{\hbar}{i}\frac{N}{x}\right)^2
+\hbar ( m\omega)=2 m (E-m^2\omega^2 x^2/2)+{ O}(1/x)\;.
\eeq
Comparing the coefficients in front of $x^2$ and $x^0$ we get $E=\hbar\omega(N+1/2)$ which is the famous formula for the spectrum of the harmonic oscillator.
In order to reconstruct
the wave function we expand \eq{ppeq} near the pole $x=x_i$. Namely, we require
\beq\la{baxterHO}
{\rm res}_{x=x_k}\left[\left(i m \omega x+\frac{\hbar}{i}\sum_{i=1}^N\frac{1}{x-x_i}\right)^2+i\hbar
\frac{\hbar}{i}\sum_{i=1}^N\frac{1}{(x-x_i)^2}\right]=0\;,
\eeq
obtaining (from the first bracket)
\beq\la{BAEHO}
x_k=\frac{\hbar}{\omega m}\sum_{j\neq i}^N\frac{1}{x_i-x_k}\;\;,\;\;k=1,\dots,N\;.
\eeq
This set of equations determines all $x_k$ in a unique way.
\begin{exercise}
Verify for $1$ and $2$ roots that there is a unique up to a permutation solution of the equation \eq{BAEHO}, find the solution.
\end{exercise}
 Finally,
we can integrating \eq{pdef} to obtain
\beq
\psi(x)=e^{-\frac{m\omega x^2}{2\hbar}}Q(x)\;\;,\;\;Q(x)\equiv\prod_{i=1}^N (x-x_i)\;.
\eeq
It is here for the first time we see the Q-function, which is the analog of the  main building block of the QSC!
We will refer to equation \eq{BAEHO} for zeros of the Q-functions
as the Bethe ansatz equation. We will call $\{x_i\}$ the Bethe roots.

Let us outline the main features which will be important for what follows:
\begin{itemize}
\item The asymptotic of  $Q(x)\sim x^N$ contains quantum numbers of the state.
\item Zeros of the $Q(x)$ function can be determined from the condition of cancellation of poles \eq{baxterHO} (analog of Baxter equation)
which can be explicitly written as \eq{BAEHO} (analog of Bethe equations).
\item The wave function can be completely determined from the Bethe roots
or from $Q(x)$ (by adding a simple universal for all states factor).
\item The Schr\"odinger equation has a second (non-normalizable)
 solution which behaves as $\psi_2\simeq x^{-N-1}e^{+\frac{m\omega}{2h}x^2}$.
 Together with the normalizable solution $\psi_1$ they form a Wronskian
 \beq
 W=\left|
 \bea{cc}\psi_1(x)& \psi_1'(x)\\
 \psi_2(x)& \psi_2'(x)
 \eea
 \right|
 \eeq
 which is a constant.
\begin{exercise}
	Prove that the Wronskian $W$ is a constant for a general Scr\"odinger equation.
\end{exercise}	
\end{itemize}

\section{$SU(2)$-Heisenberg Spin Chain}\la{sec:twist}
In this section we  discuss how the construction from the
previous section generalizes to integrable spin chains -- a system with a large number of degrees of freedom.
The simplest spin chain is the Heisenberg $SU(2)$ magnetic which
is discussed in great detail in numerous reviews and lectures.
We highly recommend Faddeev's 1982 Les Houches lectures \cite{Faddeev:1996iy} for that. We describe the results most essential for us below.

In short, the Heisenberg spin chain is a chain of $L$ spin-$1/2$ particles with a nearest neighbour interaction.
The Hamiltonian of the system can be written as
\beq\la{ham}
\hat H=2g^2\sum_{i=1}^L(1-{P}_{i,i+1})
\eeq
where $P_{i,i+1}$ is an operator which permutes the particles at the position $i$ and $i+1$ and $g$ is a constant.
We introduce twisted boundary conditions by defining 
\beqa
P_{L,L+1}|\uparrow,\dots,\uparrow\rangle=|\uparrow,\dots,\uparrow\rangle\;\;,\;\;
P_{L,L+1}|\uparrow,\dots,\downarrow\rangle=
e^{+2i\phi}|\downarrow,\dots,\uparrow\rangle\;,\\
P_{L,L+1}|\downarrow,\dots,\downarrow\rangle=|\downarrow,\dots,\downarrow\rangle
\;\;,\;\;
P_{L,L+1}|\downarrow,\dots,\uparrow\rangle=e^{-2i\phi}|\uparrow,\dots,\downarrow\rangle\;.
\eeqa
The states can, again, be described by the Baxter function $Q_1(u)=e^{\phi u}\prod_{i=1}^{N_1}(u-u_i)$.
The Bethe roots $u_i$ have a physical meaning -- they represent
the momenta $p_i$ of spin down ``excitations" moving in a sea of spin ups
via $u_i=\frac{1}{2}\cot\frac{p_i}{2}$ (see Fig.\ref{su2chain}).
We find the roots $u_j$ from the equation similar to \eq{BAEHO}\footnote{One should assume all $u_j$ to be different like in the harmonic oscillator case.}
\beq\la{BAEHO2}
\left(\frac{u_k+i/2}{u_k-i/2}\right)^L=
e^{-2i\phi}\prod_{j\neq k}^{N_1}\frac{u_k-u_j+i}{u_k-u_j-i}
\;\;,\;\;k=1,\dots,N_1\;
\eeq
\begin{exercise} 
Take log and expand for large $u_k$. You should get exactly the same as \eq{BAEHO} up to a rescaling and shift of $u_j$.
\end{exercise}
from where one gets a discrete set of solutions for $\{u_i\}$.
The energy is then given by
\beq\la{ene}
E=\sum_{j}^{N_1}\frac{2g^2}{u_j^2+1/4}\;.
\eeq

\begin{exercise}
 Take $L=2$ and compute the energy spectrum in two different ways: 1) by directly diagonalizing the Hamiltonian \eq{ham}, which becomes a $4\times 4$ matrix of the form 
$$2g^2\left(
\begin{array}{cccc}
0 & 0 & 0 & 0 \\
0 & 2 & -1-e^{-2 i \phi } & 0 \\
0 & -1-e^{2 i \phi } & 2 & 0 \\
0 & 0 & 0 & 0 \\
\end{array}
\right)$$
Next solve the Bethe equation \eq{BAEHO2} for $N_1=0,1,2$ and compute the energy from the formula \eq{ene}.
\end{exercise}

One could ask what the analog of the Schr\"odinger equation is in this case.
The answer is given by the Baxter equation of the form
\beq\la{baxsu2}
T(u)Q_1(u)=(u+i/2)^L Q_1(u-i)+(u-i/2)^L Q_1(u+i)\;,
\eeq
where $T(u)$ is a polynomial which plays the role of the potential, but it is not fixed completely and has to be determined from the self-consistency of \eq{baxsu2}.
\begin{exercise}
Show that the leading large $u$ coefficients of $T(u)$ are $T(u)\simeq 2\cos\phi u^L+
u^{L-1}(N_2-N_1)\sin\phi$ where $N_2=L-N_1$.
\end{exercise}

\noindent In practice we do not even need to know $T(u)$ as it is sufficient to require polynomiality from $T(u)$ to get \eq{BAEHO2}
as a condition of cancellation of the poles.
\begin{exercise}
For generic polynomial $Q(u)$ we see that $T(u)$ is a rational function with poles at $u=u_k$, where $Q(u_k)=0$.
Show that these poles cancel if the Bethe ansatz equation  \eq{BAEHO2} is satisfied.
\end{exercise}

\begin{figure}
\begin{center}
\def\svgwidth{\textwidth}
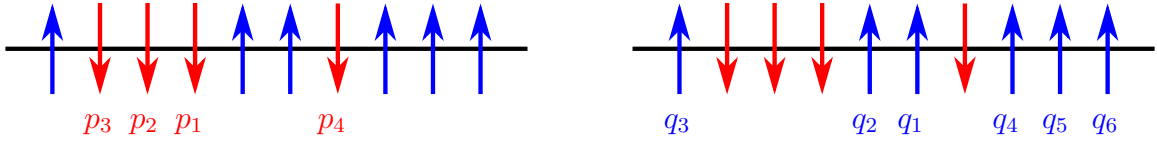
\end{center}
\caption{\la{su2chain}Two equivalent representations of the same state. In the first case
we treat spin downs as excitations (magnons) moving with some
momenta $p_i$ and all spin ups correspond to the
reference (vacuum) state. In the second case we treat spin ups as excitations moving
with some momenta $q_i$.}
\end{figure}

Notice that given some polynomial $T(u)$ there is another polynomial (up to a $e^{-u\phi}$ multiplier to ``twist" $\phi$) solution to the Baxter equation,
just like we had before for the Schr\"odinger equation.
Its asymptotics are $Q_2\simeq e^{-u\phi}u^{N_2}$ where $N_2=L-N_1$.
The roots of $Q_2$ also has a physical interpretation -- they describe the $L-N_1$ spin up particles moving in the sea of the spin downs (i.e. opposite to $Q_1$ which described the reflected picture where the spin ups played the role of the observers and the spin-downs were considered as particles).
The second solution together with the initial one should satisfy the Wronskian relation
(in the same way as for the Sch\"odinger equation)\footnote{The $\propto$ sign is used to indicate that the equality holds up to a numerical multiplier (which can be easily recovered from large $u$ limit).}
\beq\la{QQsu2}
\left|
\bea{cc}
Q_1(u-i/2)& Q_1(u+i/2)\\
Q_2(u-i/2)& Q_2(u+i/2)
\eea
\right|\propto\;Q_{12}(u)
\eeq
where $Q_{12}(u)$ satisfies 
\beq\la{Q12c}
\frac{Q_{12}(u+i/2)}{Q_{12}(u-i/2)}=\frac{(u+i/2)^L}{(u-i/2)^L}
\eeq
so we conclude that $Q_{12}(u)=-2i\sin\phi\; u^L$.
\newpage
\begin{exercise}
Show that if $Q_1$ and $Q_2$ are two linearly independent solutions of \eq{baxsu2}, then \eq{Q12c} holds.
\end{exercise}

We see that there are strict similarities with the harmonic oscillator.
Furthermore, it is possible to invert the above logic and prove the following statement:
equation \eq{QQsu2} plus the polynomiality assumption (up to an exponential prefactor)
by itself implies the Bethe equation, from which we departed.
This logic is very close to the philosophy of the QSC.

\begin{exercise}
Show that the Baxter equation is the following ``trivial" statement
\beq
\left|
\bea{lll}
Q(u-i) & Q(u)& Q(u+i)\\
Q_1(u-i) & Q_1(u)& Q_1(u+i)\\
Q_2(u-i) & Q_2(u)& Q_2(u+i)
\eea
\right|=0\;\;,\;\;{\rm for}\;\;Q=Q_1\;\;{\rm or}\;\;Q=Q_2\;.
\eeq
From that determine $T(u)$ in terms of $Q_1$ and $Q_2$.
\end{exercise} 

\section{Nested Bethe Ansatz and $QQ$-relations}

The symmetry of the Heisenberg spin chain from the previous section is $SU(2)$. In order to get closer to $PSU(2,2|4)$ (the symmetry of ${\mathcal N}=4$ SYM) we now consider a generalization of the Heisenberg
spin chain for the $SU(3)$ symmetry group. For that we just have to assume that there are $3$ possible states per chain site instead of $2$, otherwise the construction of the Hamiltonian is very similar.

The spectrum of the $SU(3)$ spin chain can be found from the ``Nested" Bethe ansatz equations~\cite{Kulish:1983rd}, which now involve two different unknown (twisted) polynomials $Q_A$ and $Q_B$. They can be written as\footnote{by the twisted polynomials we mean the functions of the form $e^{\psi u}\prod\limits_i(u-u_i)$, for some number $\psi$.}:
\beqa\la{QAB}
1&=&-\frac{Q_A^{++} Q_B^{-}}{Q_A^{--} Q_B^{+}}\;\;,\;\;u=u_{A,i}\\
\frac{Q_\theta^+}{Q_\theta^-}&=&\nn
-\frac{Q_A^- Q_B^{++}}{Q_A^+ Q_B^{--}}\;\;,\;\;u=u_{B,i}
\eeqa
and the energy is given by
\beq
E=\left.i \partial_u\log \frac{Q^+_B}{Q^-_B}\right|_{u=0}\;.
\eeq
We denote $Q_\theta=u^L$.
We also introduced some very convenient notation
\beq
f^\pm = f(u\pm i/2)\;\;,\;\;f^{\pm\pm}=f(u\pm i)\;\;,\;\;f^{[\pm a]}=f(u\pm a i/2).
\eeq
\begin{exercise}
Show that the $SU(3)$ Bethe equations reduce to the $SU(2)$ equations
\eq{BAEHO2} and \eq{ene} when $Q_A=1$.
\end{exercise}

\subsection{Bosonic duality}
From the $SU(2)$ Heisenberg spin chain we learned that the Baxter polynomial $Q_1(u)$ contains as many roots as arrow-downs we have in our state. In particular the trivial polynomial $Q_1(u)=e^{-u\phi}$ corresponds to the state $|\uparrow\uparrow\dots\uparrow\rangle
$. One can also check that there is only one solution of the Bethe equations where $Q_1(u)$ is a twisted polynomial of degree $L$ and it satisfies
\beq
e^{i\phi/2}Q_1^--e^{-i\phi/2}Q_1^+=2i\sin\phi u^Le^{-u\phi} \; .
\eeq
\begin{exercise}
Solve this equation for $L=1$ and $L=2$ and check that
$Q_1$ also solves the Bethe equations of the $SU(2)$ spin chain. Compute the corresponding energy.
\end{exercise}
As this equation produces a polynomial of degree $L$ it must correspond to the maximally ``excited" state $|\downarrow\downarrow\dots\downarrow\rangle$.
It is clear that even though physically these states are very similar our current description in terms of the Bethe ansatz singles out one of them.
We will see that there is a ``dual" description where the Q-function corresponding to the state
$|\downarrow\downarrow\dots\downarrow\rangle$ is trivial. In the case of the $SU(3)$ spin chain where we have $3$ different states per node of the spin chain, which we can denote $1,2,3$, there are $3$ equivalent vacuum states $|11\dots 1\rangle$, $|22\dots 2\rangle$, and $|33\dots 3\rangle$, but only one of them corresponds to the trivial solution of the $SU(3)$ nested Bethe ansatz. Below we concentrate on the $SU(3)$ case and demonstrate that there are several equivalent sets of Bethe ansatz equations \eq{QAB}.

\begin{figure}
\begin{center}
\def\svgwidth{0.5\textwidth}
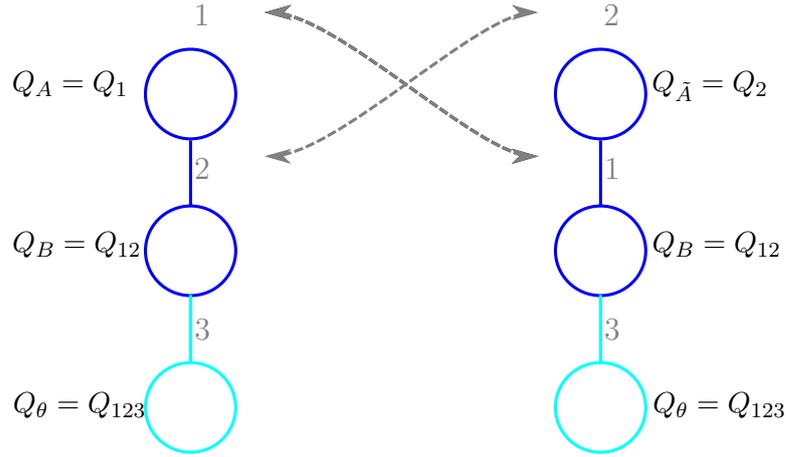
\end{center}
\caption{\la{su2chain2}Bosonic duality applied to the first node of the BA.}
\end{figure}

To build a dual set of Bethe equations we first have to pick a $Q$-function which we are going to dualise. For example we can build a new set of Bethe equations by replacing $Q_{A}$, a twisted polynomial of degree $N_A$,
with another twisted polynomial $Q_{\tilde A}$ of degree $N_{\tilde A}=N_B-N_A$, where $N_B$ is the degree of the polynomial $Q_B$. 
For that we find a dual $Q$-function $Q_{\tilde A}$ from
\beq\la{Bdualitysu2}
\left|
\bea{cc}
Q_A^-& Q_A^+\\
Q_{\tilde A}^-& Q_{\tilde A}^+
\eea
\right|\propto\;Q_{B}(u)\;.
\eeq
Let's see that $Q_{\tilde A}$ satisfies the same Bethe equation.
By evaluating \eq{Bdualitysu2} at $u=u_{\tilde A,i}+i/2$
and dividing by the same relation evaluated at $u=u_{\tilde A,i}-i/2$ we get:
\beq
\frac{Q_A Q_{\tilde A}^{++}-0}
{0-Q_A Q_{\tilde A}^{--}}=\frac{Q_B^+}{Q_B^-}\;\;,\;\;u=u_{\tilde A,i}
\eeq
which is exactly the first equation \eq{QAB} with $A$ replaced by $\tilde A$!
To accomplish our goal we should also exclude $Q_A$ from the second equation. For that we notice that at $u=u_{B,i}$
the relation gives
\beq
\frac{Q_A^-}{Q_A^+}=\frac{Q_{\tilde A}^-}{Q_{\tilde A}^+}\;\;,\;\;u=u_{B,i}\;
\eeq
which allows us to rewrite the whole set of equations 
\eq{QAB} in terms of $Q_{\tilde A}$.
We call this transformation a Bosonic duality. Similarly one can 
apply the dualization procedure to $Q_B$. We determine $Q_{\tilde B}$
from
\beq\la{Bdualitysu2_2}
\left|
\bea{cc}
Q_B^-& Q_B^+\\
Q_{\tilde B}^-& Q_{\tilde B}^+
\eea
\right|\propto\;Q_{A}(u)Q_{\theta}(u)\;.
\eeq

By doing this we will be able to replace $B$ by $\tilde B$ in \eq{QAB}.
Let us also show that we can use $Q_{\tilde B}$ instead of $Q_{B}$ in the expression for the energy \eq{ene}. We recall that $Q_\theta(u)\propto u^L$, so evaluating \eq{Bdualitysu2_2} at $u=0$ we get
\beq\la{QQ0}
Q_B(-\tfrac{i}{2})Q_{\tilde B}(+\tfrac{i}{2})=Q_B(+\tfrac{i}{2})Q_{\tilde B}(-\tfrac{i}{2})\;.
\eeq
We can also differentiate \eq{Bdualitysu2_2} in $u$ once and then set $u=0$, so that
\beq\la{dirQ}
Q_B'(-\tfrac{i}{2})Q_{\tilde B}(+\tfrac{i}{2})+Q_B(-\tfrac{i}{2})Q'_{\tilde B}(+\tfrac{i}{2})=Q'_B(+\tfrac{i}{2})Q_{\tilde B}(-\tfrac{i}{2})+Q_B(+\tfrac{i}{2})Q'_{\tilde B}(-\tfrac{i}{2})\;.
\eeq
Dividing \eq{dirQ} by \eq{QQ0} we get
\beq
\frac{Q_B'(-\tfrac{i}{2})}{Q_{ B}(-\tfrac{i}{2})}+\frac{Q'_{\tilde B}(+\tfrac{i}{2})}{Q_{\tilde B}(+\tfrac{i}{2})}=
\frac{Q'_B(+\tfrac{i}{2})}{Q_B(+\tfrac{i}{2})}+\frac{Q'_{\tilde B}(-\tfrac{i}{2})}{Q_{\tilde B}(-\tfrac{i}{2})}\;,
\eeq
which indeed gives
\beq
E=\left.i \partial_u\log \frac{Q^+_B}{Q^-_B}\right|_{u=0}=\left.i \partial_u\log \frac{Q^+_{\tilde B}}{Q^-_{\tilde B}}\right|_{u=0}\;.
\eeq
\paragraph*{Better notation for $Q$-functions}
One can combine the above duality transformations and say dualise $Q_A$
after dualising $Q_B$ and so on. In order to keep track of all possible transformations one should introduce some notation, as otherwise we can end up with multiple tildas.
Another question we will try to answer in this part is how many  equivalent BA's we will generate by applying the duality many times to various nodes.

In order to keep track of the dualities we place numbers $1,2,3$ in between the nodes of the Dynkin diagram.
We place the $Q$-functions on the nodes of the diagram as in Fig.\ref{su2chain2}.
Then we interpret the duality as an exchange of the corresponding labels sitting on the links of the diagram, so if before the dualization of $Q_A$ we had $1,2,3$, after the duality we have to exchange the indexes $1$ and $2$ obtaining $2,1,3$. If instead we first dualised $Q_B$ we would obtain $1,3,2$. Each duality produces a permutation of the numbers.
We also use these numbers to label the $Q$-functions. Namely we assign the indexes to the $Q$ function in accordance with the numbers appearing above the given node. So, in particular, in the new notation
\beq
Q_{A}=Q_1\;\;,\;\;Q_{B}=Q_{12}\;.
\eeq
Each order of the indexes naturally corresponds to a particular set of Bethe equations. For instance, the initial set of Bethe equations on $Q_{A}, Q_{B}$
correspond to the order $1,2,3$ and the Bethe ansatz (BA) for $Q_{\tilde A}, Q_{B}$ correspond to $2,1,3$ and so on. Now we can answer the question of how many dual BA systems we could have;
this is given by the number of permutations of $1,2,3$ i.e. for the case of $SU(3)$ we get $6$ equivalent systems of BA equations.

Following our prescription we also denote
\beq
Q_{\tilde A}=Q_2\;\;,\;\;Q_{\tilde B}=Q_{13}\;.
\eeq
We note that we should not distinguish Q's which only differ by the order of indexes. So, for instance, $Q_{21}$ and $Q_{12}$ is the same $Q_{B}$. 
We can count the total number of various $Q$-functions we could possibly generate with the dualities: $2^3-2=6$ different $Q$-functions
which are
\beq
Q_{i}\;\;,\;\;Q_{[ij]}\;\;,\;\;i,j=1,\dots,3
\eeq
for completeness we also add $Q_{\emptyset}\equiv 1$ and $Q_{123}=Q_{[ijk]}\equiv Q_\theta=u^L$ so that in total we have $2^3$. For general $SU(N)$ we will find $2^N$ different Q-functions.
We see that the number of the Q-functions grows rapidly with the rank of the symmetry group.
For  $PSU(2,2|4)$ we get $256$ functions, and we should study the relations among them in more detail.
\paragraph*{$QQ$-relations}
Let us rewrite the Bosonic duality in the new notation.
The relation \eq{Bdualitysu2} becomes
\beq\la{QQ2}
\left|
\bea{cc}
Q_i^-& Q_i^+\\
Q_j^-& Q_j^+
\eea
\right|\propto\;Q_{ij}Q_{\emptyset}
\eeq
where we added $Q_\emptyset=1$ to the r.h.s. to make both l.h.s and r.h.s be bilinear in $Q$. Very similarly \eq{Bdualitysu2_2} gives
\begin{eqnarray}\la{qqq1}
\left|
\bea{cc}
Q_{1{\bf 2}}^-& Q_{1{\bf 2}}^+\\
Q_{1{\bf 3}}^-& Q_{1{\bf 3}}^+
\eea
\right|\propto\;Q_{1}(u)Q_{1{\bf 23}}(u)\;.
\end{eqnarray}
We see that both identities can be written in one go as
\begin{eqnarray}\la{qqq2}
\left|
\bea{cc}
Q_{I{\bf i}}^-& Q_{I{\bf i}}^+\\
Q_{I{\bf j}}^-& Q_{I{\bf j}}^+
\eea
\right|\propto\;Q_{I}(u)Q_{I{\bf ij}}(u)\;,
\end{eqnarray}
where for general $SU(N)$ we would have $i=1,\dots,N$
and $j=1,\dots,N$ and $I$ represents a set of indexes
such that in \eq{qqq1} it is an empty set $I=\emptyset$
and for the second identity \eq{qqq2}  $I$ contains only one element $1$.
Note that no indexes inside $I$ are involved with the relations and
in the r.h.s. we get indexes $i$ and $j$ glued together in the new function. We see that proceeding in this way we can build any $Q$-function starting from the basic $Q_i$ with one index only. For that we can first take $I=\emptyset$ and build $Q_{ij}$, then take $I=i$
and build $Q_{ijk}$ and so on. It is possible to combine these steps together to get explicitly
\beq\la{QQ3}
Q_{ijk}Q^+_\emptyset Q^-_\emptyset\propto\left|
\bea{ccc}
Q_i^{--}& Q_i& Q_i^{++}\\
Q_j^{--}& Q_j& Q_j^{++}\\
Q_k^{--}& Q_k& Q_k^{++}
\eea
\right|\;\;.
\eeq
Whereas the first identity \eq{QQ2} is obvious from the definition, the second \eq{QQ3} is a simple exercise to prove from
\eq{qqq2}.
\begin{exercise}
Prove \eq{QQ3} using the following Mathematica code
\begin{mathematica}
(*define Q to be absolutely antisymmetric*)
Q[a___] := Signature[{a}] Q @@ Sort[{a}] /; ! OrderedQ[{a}]
(*program bosonic duality*)
Bosonic[J___, i_, j_] := Q[J, i, j][u_] -> ( 
Q[J, i][u + I/2] Q[J, j][u - I/2] - 
Q[J, i][u - I/2] Q[J, j][u + I/2])/Q[J][u];
(*checking the identity*)
Q[1, 2, 3][u] Q[][u + I/2] Q[][u - I/2] /. Bosonic[1, 2, 3] /. 
Bosonic[1, 2] /. Bosonic[1, 3] /. Bosonic[2, 3] // Factor
\end{mathematica}
Also derive a similar identity for $Q_{ijkl}$ using the same code.
\end{exercise}

From the previous exercise it should be clear that we can generate any $Q_{ij\dots k}$ as a determinant of the
basic $N$ Q-functions $Q_i$. In particular the ``full-set" Q-function $Q_{12\dots N}$, which is also $Q_\theta=u^L$, can be written as a determinant of $N$ basic polynomials $Q_i$. Interestingly this identity by itself is constraining enough to give rise to the full spectrum of the $SU(N)$ spin chain! Indeed $Q_{12\dots N}$ is a polynomial of degree $L$ and thus we get $L$ nontrivial relations
on the coefficients of the (twisted) polynomials $Q_i$, which together contain exactly $L$ Bethe roots. This means that this relation alone is equivalent to the whole set of Nested Bethe ansatz equations.
So we can put aside a non-unique BA approach, dependent on the choice of the vacuum, and replace it completely by a simple determinant like \eq{QQ3}. In other words the QQ-relations and the condition of polynomiality is all we need to quantize this quantum integrable model.
We will argue that for ${ N}=4$ SYM we only have to replace the polynimiality with another slightly more complicated analyticity condition but otherwise keep the same QQ-relations. We will have to, however, understand what the QQ-relations look like for the case of super-symmetries like $SU(N|M)$, which is described in the next section. 

\subsection{Fermionic duality in $SU(N|M)$}

We will see how the discussion in the previous section generalizes to the super-group case. Our starting point will be again the set of nested Bethe ansatz equations, which follow the pattern of the Cartan matrix.
Let us discuss the construction of the Bethe ansatz. Below we wrote the Dynkin diagram, Cartan matrix and the Bethe ansatz equations for the $SU(3|3)$ super spin chain
\begin{eqnarray}\la{BAEs}
\bea{ccc}
\bea{cc}
Q_A&\bigcirc\\
Q_B&\bigcirc\\
Q_C&\bigotimes\\
Q_D&\bigcirc\\
Q_E&\bigcirc
\eea &
\bea{|c|c|c|c|c|}
\hline
2&-1&0&0&0\\ \hline
-1&2&-1&0&0\\ \hline
0&-1&0&+1&0\\ \hline
0&0&+1&-2&+1\\ \hline
0&0&0&+1&-2\\ \hline
\eea &
\bea{l}
-1=(\;\;\;\;\;Q^{++}_A Q_B^-)/(\;\;\;\;\;Q^{--}_A Q_B^+)\;\;,\;\;u=u_{A,i}\\
-1=(Q_A^-Q^{++}_B Q_C^-)/(Q_A^+Q^{--}_B Q_C^+)\;\;,\;\;u=u_{B,i}\\
+1=(Q^{-}_B\;\;\;\;\;\;\;Q_D^+)/(Q^{+}_B\;\;\;\;\;\;\,Q_D^-)\;\;,\;\;u=u_{C,i}\\
-1=(Q_C^+Q^{--}_D Q_E^+)/(Q_C^-Q^{++}_D Q_E^-)\;\;,\;\;u=u_{D,i}\\
-1=(Q_D^+Q^{--}_E \;\;\;\;\;)/(Q_D^-Q^{++}_E\;\;\;\;\;)\;\;,\;\;u=u_{E,i}\\
\eea
\eea
\end{eqnarray}
The $Q$-functions still correspond to the nodes of the Dynkin diagrams
and the shift of the argument of the $Q$-functions entering the numerators of the Bethe equations simply follow the pattern of the Cartan matrix (with the inverse shifts in numerators).
Since the structure of the equations for the bosonic  nodes is the same as before, one can still apply the Bosonic duality transformation for instance on $Q_B$ and replace it by $Q_{\tilde B}$. However for the fermionic type nodes (normally denoted by a crossed circle), such as $Q_C$, we get a new type of duality transformation
\beq\la{Qfdual}
Q_C Q_{\tilde C}\propto
\left|
\bea{cc}
Q_B^-& Q_B^+\\
Q_D^-& Q_D^+
\eea
\right|
\eeq
which look similar to the Bosonic one with the difference that we can extract explicitly the dual Baxter polynomial $Q_{\tilde C}$\footnote{Whereas for the Bosonic duality \eq{Bdualitysu2} the dual Baxter polynomial occur in a complicated way and one had to solve a first order finite difference equation in order to extract it.}.
Let us show that the middle Bethe equation can be obtained from the duality relation \eq{Qfdual}.
Indeed we see again that for both $u=u_{C,i}$ and $u=u_{\tilde C,i}$
we get the middle equation 
\beq
1=\frac{Q^+_{B}Q^-_{D}}{Q^-_{B}Q^+_{D}}\;\;,\;\;u=u_{\tilde C,i}\;\;{\rm or}\;\;u=u_{C,i}.
\eeq
Next we should be able to exclude $Q_C$ in the other two equations. For that we set $u=u_{B,i}+i/2$ and $u=u_{B,i}+i/2$ to get
\beq
Q^+_CQ^+_{\tilde C}=c(0- Q_D Q_B^{++})\;\;,\;\;
Q^-_CQ^-_{\tilde C}=+c(Q_D Q_B^{--}-0)\;\;u=u_{B,i}\;.
\eeq
Dividing one by the other
\beq
-1=\frac{Q^+_CQ^+_{\tilde C}}
{Q^-_CQ^-_{\tilde C}}\frac{Q_B^{--}}{Q_B^{++}}\;\;,\;\;u=u_{B,i}
\eeq 
which allows up to exclude $Q_C$ from the second equation of \eq{BAEs}. This then becomes
\beq
-1=\frac{Q_A^-Q^{++}_B Q_C^-}{Q_A^+Q^{--}_B Q_C^+}\;\;\leftrightarrow\;\;
+1=\frac{Q_A^-Q_{\tilde C}^+}{Q_A^+Q_{\tilde C}^-}
\;\;,\;\;u=u_{B,i}\;.
\eeq
As we see this changes the type of the equation from bosonic to fermionic. Thus we also change the type of the Dynkin diagram. This is expected since for super algebras the Dynkin diagram is not unique. Similarly the fourth equation also changes in a similar way. To summarize,
after duality we get 
\begin{eqnarray}\la{BAEs}
\bea{ccc}
\bea{cc}
Q_A&\bigcirc\\
Q_B&\bigotimes\\
Q_{\tilde C}&\bigotimes\\
Q_D&\bigotimes\\
Q_E&\bigcirc
\eea &
\bea{|c|c|c|c|c|}
\hline
2&-1&0&0&0\\ \hline
-1&0&+1&0&0\\ \hline
0&+1&0&-1&0\\ \hline
0&0&-1&0&+1\\ \hline
0&0&0&+1&-2\\ \hline
\eea &
\bea{l}
-1=(\;\;\;\;\;Q^{++}_A Q_B^-)/(\;\;\;\;\;Q^{--}_A Q_B^+)\;\;,\;\;u=u_{A,i}\\
+1=(Q_A^-\;\;\;\;\;\;\; Q_{\tilde C}^+)/(Q_A^+\;\;\;\;\;\;\; Q_{\tilde C}^-)\;\;,\;\;u=u_{B,i}\\
+1=(Q^{+}_B\;\;\;\;\;\;\;Q_D^-)/(Q^{-}_B\;\;\;\;\;\;\,Q_D^+)\;\;,\;\;u=u_{\tilde C,i}\\
+1=(Q_{\tilde C}^-\;\;\;\;\;\;\;Q_E^+)/(Q_{\tilde C}^+\;\;\;\;\;\;\; Q_E^-)\;\;,\;\;u=u_{D,i}\\
-1=(Q_D^+Q^{--}_E \;\;\;\;\;)/(Q_D^-Q^{++}_E\;\;\;\;\;)\;\;,\;\;u=u_{E,i}\\
\eea
\eea
\end{eqnarray}
\begin{figure}
	\begin{center}
		\def\svgwidth{0.5\textwidth}
		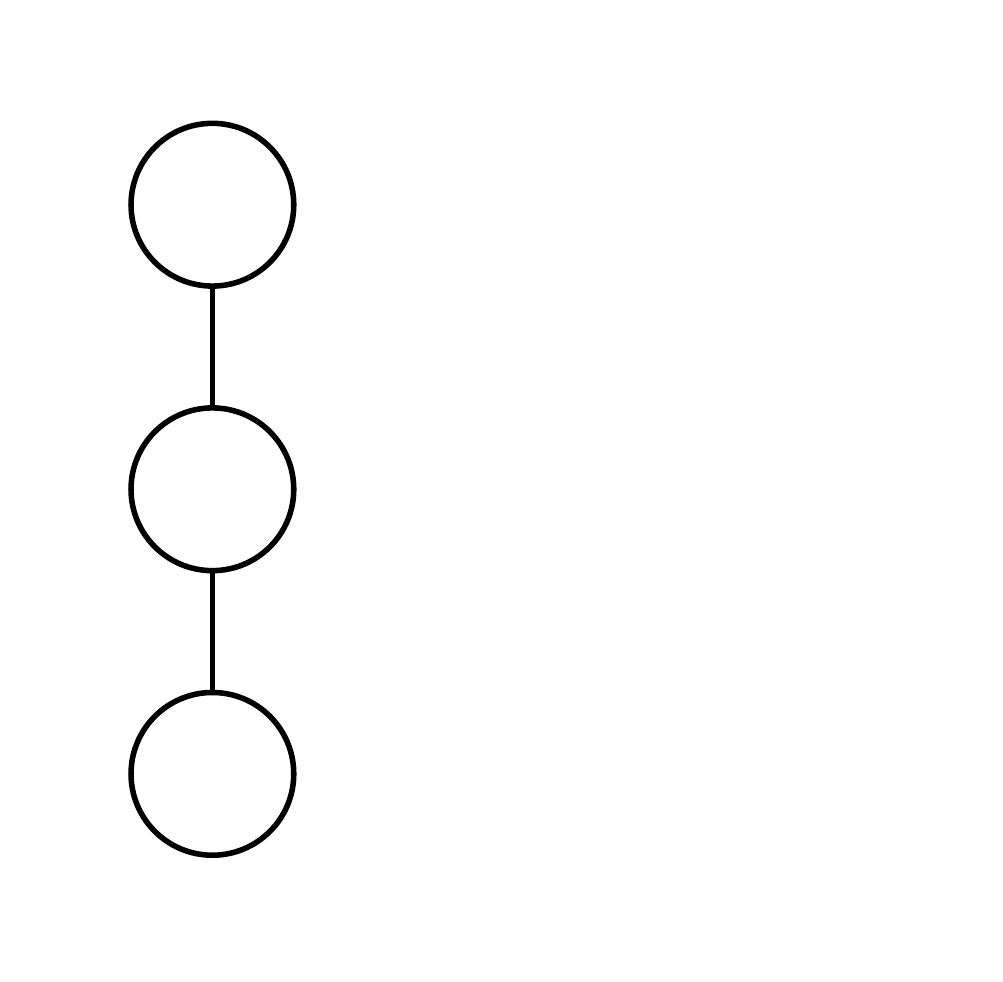
	\end{center}
	\caption{\la{su2chainf}Fermionic duality}
\end{figure}

\paragraph*{Index notation}
Again in order to keep track of all possible combinations of dualities we have to introduce index notation.
In the super case we label the links in the Dynkin diagram by two types of indexes (with hat and without). The type of the index changes each time we cross a fermionic node. For instance our initial set of Bethe equations corresponds to the indexes $123\hat 1\hat 2\hat 3$. The fermionic duality transformation again simply exchanges the labels on the links of the Dynkin diagram (see Fig.\ref{su2chainf}). So after duality we get $12\hat 1 3\hat 2\hat3$, which is consistent with the $\bigcirc-\bigotimes-\bigotimes-\bigotimes-\bigcirc$ grading of the resulting Bethe ansatz equations.
Finally, we label the $Q$-functions by two antisymmetric groups of indexes -- with hat and without again simply listing all indexes appearing above the given node of the Dynkin diagram. In particular we get
\beqa
&Q_A=Q_{1}\;\;,\;\;Q_B=Q_{12}\;\;,\;\;Q_C=Q_{123}\;\;,&\\
\nn&Q_{\tilde C}=Q_{12\hat 1}\;\;,\;\;Q_D=Q_{123\hat 1}\;\;,\;\;Q_E=Q_{123\hat 1\hat 2}\;.&
\eeqa
An alternative notation is to omit hats and separate the two sets of indexes by a vertical line: 
\beqa
&Q_A=Q_{1|\emptyset}\;\;,\;\;Q_B=Q_{12|\emptyset}\;\;,\;\;Q_C=Q_{123|\emptyset}\;\;,&\\
\nn&Q_{\tilde C}=Q_{12|1}\;\;,\;\;Q_D=Q_{123| 1}\;\;,\;\;Q_E=Q_{123|12}\;.&
\eeqa
\begin{exercise}
The fermionic duality transformation changes the type of the Dynkin diagram. The simplest way to understand which diagram one gets after the duality is to follow the indexes attached to the links. Each time the type of the index changes (from hatted to non-hatted)
you should draw a cross. List all possible Dynkin diagrams corresponding to $SU(3|3)$ Lie algebra.
\end{exercise}

\paragraph*{Fermionic QQ-relations}
In index notation \eq{Qfdual} becomes
\beq\la{fe}
\boxed{
Q_{I b} Q_{I \hat i}\propto
Q_{I}^-Q_{I b\hat i}^+
-
Q_{I}^+Q_{I b\hat i}^-
}\;.
\eeq
For completeness let us write here the bosonic duality relations
\beq\la{bo}
\boxed{
Q_{I}Q_{I a b}\propto
Q_{I a}^+Q_{I b}^-
-
Q_{I b}^+Q_{I a}^- \;\;,\;\;
Q_{I}Q_{I \hat i\hat j}\propto
Q_{I \hat i}^+Q_{I \hat j}^-
-
Q_{I \hat j}^+Q_{I \hat i}^- 
}\;.
\eeq

In the case of $SU(N|M)$ one could derive all $Q$ functions in terms of $N+M$ functions $Q_a$
and $Q_{\hat i}$. We will demonstrate this in the next section in the example of $SU(4|4)$.

\section{QQ-relations for $PSU(2,2|4)$ Spin Chain}\la{sec:PSU224QQ}
The global symmetry of ${\mathcal N}=4$ SYM is $PSU(2,2|4)$. The QQ-relations from the previous section associated with this symmetry constitute an important part of the QSC construction.
The symmetry (up to a real form and a projection) is same as $SU(4|4)$.
In this section we specialize the QQ-relations from the previous part to this case and derive all the most important relations among Q-functions.
In particular we show that all $256$ various Q-functions can be derived from just $4+4$ Q-functions with one index
\beq
Q_{a|\emptyset}\;\;,\;\;Q_{\emptyset|i}\;,
\eeq
which are traditionally denoted in the literature as
\beq
{\bf P}_a\;\;,\;\;{\bf Q}_i\;.
\eeq
These are the elementary Q-functions.

For us, another important object is $Q_{a|i}$. According to the general consideration above it
can be obtained from the fermionic duality relation \eq{fe} with $I=\emptyset$
\beq\la{Qai}
Q_{a| j}^+-Q_{a| j}^-={\bf P}_a{\bf Q}_j\;.
\eeq
This is the first order equation on $Q_{a|i}$ which one should solve; and the formal solution to this equation is\footnote{Note that there is a freedom to add a constant to $Q_{a|i}$. This freedom is fixed in the twisted case as we should require that $Q_{a|j}$ has a ``pure" asymptotics at large $u$ i.e. $e^{\phi_{ai}u}u^\alpha(1+A_1/u+A_2/u^2+\dots)$.}
\beq\la{Qaiformal}
Q_{a|j}(u)=-\sum_{n=0}^\infty {\bf P}_a(u+i\tfrac{2n+1}{2}){\bf Q}_i(u+i\tfrac{2n+1}{2})\;.
\eeq
\begin{exercise}
Find a solution to the equation \eq{Qai}
for ${\bf P}_a{\bf Q}_j=e^{\phi u}$
and also for ${\bf P}_a{\bf Q}_j=1/u^2$.
\end{exercise}

Once we know $Q_{a|i}$ we can build any $Q$-function explicitly in terms of $Q_{a|i},\;\bQ_i$ and $\bP_a$. For example using the Bosonic duality we can get
\beq
Q_{{\bf ab}|i}=\frac{Q_{{\bf a}|i}^+Q_{{\bf b}|i}^--Q_{{\bf a}|i}^-Q_{{\bf b}|i}^+}{\bQ_i}\;.
\eeq
In this way we can build all Q-functions explicitly in terms of $Q_{a|i},\;\bQ_i$ and $\bP_a$. There is a nice simplification
taking place for Q-functions with equal number of indexes:
\beq\la{Q22}
Q_{ab|ij}=\left|
\bea{cc}
Q_{a|i}&Q_{a|j}\\
Q_{b|i}&Q_{b|j}
\eea
\right|
\eeq
\begin{exercise}
Prove \eq{Q22} using the following Mathematica code
\begin{mathematica}
(*define Q to be absolutely antisymmetric*)
Q[a___][b___][u_] := Signature[{a}] Signature[{b}] 
Q[Sequence @@ Sort[{a}]][Sequence @@ Sort[{b}]][u] 
/; ! (OrderedQ[{a}] && OrderedQ[{b}])
(*program bosonic and fermionic dualities*)
B1[J___, a_, b_][K___] := Q[J, a, b][K][u_] :> 
(Q[J, a][K][u + I/2] Q[J, b][K][u - I/2] - 
Q[J, a][K][u - I/2] Q[J, b][K][u + I/2])/Q[J][K][u];
B2[K___][J___, i_, j_] := 
Q[K][J, i, j][u_] :> (Q[K][J, i][u + I/2] Q[K][J, j][u - I/2] - 
Q[K][J, i][u - I/2] Q[K][J, j][u + I/2])/Q[K][J][u];
F1[K___, a_][J___, i_][u_] := Q[K, a][J, i][u] :> 
(Q[K, a][J, i][u - I] Q[K][J][u - I] + 
Q[K, a][J][u - I/2] Q[K][J, i][u - I/2])/Q[K][J][u]
F2[K___, a_][J___, i_][u_] := Q[K, a][J, i][u] :> 
(Q[K, a][J, i][u + I] Q[K][J][u + I] - 
Q[K, a][J][u + I/2] Q[K][J, i][u + I/2])/Q[K][J][u]
(*deriving the identity*)
Q[a, b][i, j][u] /. B1[a, b][i, j] /. B2[a][i, j] /. B2[b][i, j] /. 
Flatten[Table[F1[c][k][u + I], {c, {a, b}}, {k, {i, j}}]] /. 
Flatten[Table[F2[c][k][u - I], {c, {a, b}}, {k, {i, j}}]] /. 
B2[][i, j] // Simplify
\end{mathematica}
Also derive a similar identity for $Q_{abc|ijk}$ using the same code.
The general strategy is to use the bosonic duality to decompose $Q$'s into $Q$-functions with fewer indexes. Then use \eq{Qai} to bring all $Q_{a|k}(u+i n)$ to the same argument $Q_{a|k}(u)$.
After that the expression should simplify enormously.
Also show the following identities to hold
\begin{eqnarray}\la{Q3n4}
Q_{abc|ijkl}=
\bQ_i Q^+_{abc|jkl}-
\bQ_j Q^+_{abc|kli}+
\bQ_k Q^+_{abc|lij}-
\bQ_l Q^+_{abc|ijk}\;,\\
Q_{abcd|ijk}=\la{Q4n3}
\bP_a Q^+_{bcd|ijk}-
\bP_b Q^+_{cda|ijk}+
\bP_c Q^+_{dab|ijk}-
\bP_d Q^+_{abc|ijk}\;.
\end{eqnarray}
Also check \eq{Q12341234} and \eq{periodicity} below.
\end{exercise}
In particular for the $Q$-function with all indexes $Q_{1234|1234}$ (remember that the $Q$-function with all indexes played an important role in the XXX spin chain giving the external ``potential" $Q_\theta=u^L$)
we get
\beq\la{Q12341234}
Q_{1234| 1 2 3 4}=
\left|
\bea{cccc}
Q_{1|1}&Q_{1|2}&Q_{1|3}&Q_{1|4}\\
Q_{2|1}&Q_{2|2}&Q_{2|3}&Q_{2|4}\\
Q_{3|1}&Q_{3|2}&Q_{3|3}&Q_{3|4}\\
Q_{4|1}&Q_{4|2}&Q_{4|3}&Q_{4|4}
\eea
\right|\;.
\eeq
Finally, one can show that
\beq\la{periodicity}
Q_{1234|1234}^{+}-Q_{1234|1234}^-=\sum_{a,i}\bQ_i\bP_a Q^-_{1234\check a,1234 \check i}
\eeq
where the check (inverse hat) denotes an ``index annihilator" i.e. for example $Q_{1234\check 4|\dots}=Q_{123|\dots}$ and $Q_{1234\check 3|\dots}=-Q_{123\check 3 4|\dots}=-Q_{124|\dots}$ and so on.

\paragraph*{Hodge duality} The $SU(4|4)$ Dynkin diagram has an obvious symmetry -- we can flip it upside down. At the same time the labeling of the $Q$-functions essentially breaks this symmetry as we agreed to list all indexes from above a given node and not below. To fix this we can introduce a Hodge dual set of $Q$-functions by defining
\begin{eqnarray}\la{hdu}
Q^{a_1\dots a_n|i_1\dots i_m}\equiv (-1)^{n m} \epsilon^{a_1\dots a_n b_1\dots b_{4-n}}\epsilon^{i_1\dots u_m j_1\dots j_{4-m}}Q_{b_1\dots b_{4-n}|j_1\dots j_{4-m}}
\end{eqnarray} 
with $b_1<\dots<b_{4-n}$ and $j_1<\dots<j_{4-n}$ so that there is only one term in the r.h.s.
One can check that these $Q$-functions with upper indexes satisfy the same QQ-relations as the initial $Q$-functions\footnote{in particular \eq{hdu} implies $Q^{\emptyset|1}=+Q_{1234|234}$ and $Q^{\emptyset|2}=-Q_{1234|134}$ and so on.
}.

Finally, we already set $Q_{\emptyset|\emptyset}=1$ and considering the symmetry of the system we should also set $Q_{1234|1234}=Q^{\emptyset|\emptyset}=1$. In fact that is indeed the case for ${\mathcal N}=4$ SYM whereas for the spin chains we have $Q_\theta=u^L$ attached to one of the ends of the Dynkin diagram, which breaks the symmetry.

Assuming $Q_{1234|1234}=1$ we get some interesting consequences. In particular the l.h.s. of \eq{periodicity} vanishes and we get
\beq\la{zero}
\bQ_i\bP_aQ^{a|i}=0\;.
\eeq
Also we can rewrite \eq{Q3n4} and \eq{Q4n3} in our new notation
\beqa
\bP^a&\equiv& Q^{a|\emptyset}=Q^{a|i}(u+i/2)\bQ_i\;,\\
\bQ^i&\equiv& Q^{\emptyset|i}=Q^{a|i}(u+i/2)\bP_a\;.
\eeqa
Combining that with \eq{zero} we get
\beq
\bP_a\bP^a=\bQ_i\bQ^i=0\;.
\eeq
Finally we can expand the determinant of the $4\times 4$ matrix in \eq{Q12341234} in the first row to get
\beq
1=Q_{1|1}Q_{234|234}
-Q_{1|2}Q_{234|134}
+Q_{1|3}Q_{234|124}
-Q_{1|4}Q_{234|123}\;,
\eeq
which is equivalent to $-1=Q_{1|a}Q^{1|a}$. Also we can replace the first row in \eq{Q12341234} by $Q_{2|i}$ instead of $Q_{1|i}$ to get zero determinant. At the same time expanding this determinant in the first row will result in $0=Q_{2|a}Q^{1|a}$. At the end we will get the following general expression
\beq\la{QupQdn}
Q_{i|a}Q^{j|a}=-\delta_i^j\;
\eeq
which implies that $Q^{i|a}$ is inverse to $Q_{i|a}$.

With these relations, we have completed the task of building the QQ-relations for $SU(4|4)$ symmetry (with an additional condition that $Q_{1234|1234}=1$, which can be associated with `P' in $PSU(2,2|4)$). The next step is to understand the analytical  properties of the $Q$-functions.
For the case of the spin chain all $Q$-functions are simply polynomials and it was sufficient to produce the spectrum from the QQ-relations. However, in that construction there is no room for a continuous parameter -- the 't Hooft coupling $g=\frac{\sqrt\lambda}{4\pi}$ and
thus for  the ${ N}=4$ SYM the analytical properties should be more complicated and we will motivate the analyticity in the next section.
The analytical properties are the missing ingredients in the construction and to deduce them we will have to revise the
strong coupling limit.

\chapter{Classical String and Strong Coupling Limit of QSC}\la{sec:classics}\la{ch:3}
In this section we briefly describe the action of the super-string in $AdS^5\times S_5$, following closely \cite{Gromov:2007aq}.
We also advice to study the lecture notes of K.Zarembo from the same Les Houches summer school.
\section{Classical String Action}
The classical action is similar to the action of the principal chiral field (PCF), so let us briefly review it. The fields $g(\sigma,\tau)$ in PCF belong to the $SU(N)$ group. One builds ``currents" out of them by
\beq\la{current}
J_\mu\equiv -g^{-1} \partial_\mu g
\eeq 
and then the classical action is simply
\beq
S=\frac{\sqrt{\lambda}}{4\pi}\int{\rm tr} (J\wedge J)\;.
\eeq
The global symmetry of this action is $SU_L(N)\times SU_R(N)$ since we can change $g(\sigma,\tau)\to h_L g(\sigma,\tau) h_R$ for arbitrary $h_L,h_R\in SU(N)$ without changing the action.

The construction for the Green-Schwartz superstring action is very similar. We take $g\in SU(2,2|4)$ and then the current $J$ (taking values in the $su(2,2|4)$ algebra) is built in the same way as in \eq{current}.
The only new ingredient is that we have to decompose the current into $4$ components in order to ensure an extra local $sp(2,2)\times sp(4)$ symmetry in the way described below.

The superalgebra $su(2,2|4)$ can be represented by $8\times 8$ supertraceless supermatrices
\beq
M=\left(
\bea{c|c}
A&B\\ \hline
C&D
\eea
\right)
\eeq 
where $A\in u(2,2)$ and $B\in u(4)$ and the fermionic components are related by 
\beq
C=B^\dagger\left(
\bea{cc}1_{2\times 2}&0\\
0&-1_{2\times 2}
\eea
\right)\;.
\eeq
An important property of the $su(2,2|4)$ superalgebra is that there is a $Z_4$ automorphism (meaning that one should act $4$ times to get a trivial transformation). This $Z_4$ automorphism has its counterpart in the QSC construction as we discuss later. Its action on an element of the algebra is defined in the following way:
\beq
\phi[M]\equiv \left(
\bea{c|c}
E A^T E &-E C^T E\\ \hline
E B^T E & E D^T E
\eea
\right)\;\;,\;\;E=\left(
\bea{cccc}
0&-1&0&0\\
1&0&0&0\\
0&0&0&-1\\
0&0&1&0
\eea
\right)\;.
\eeq
It is easy to see that $\phi^4=1$. The consequence of this is that any element of the algebra can be decomposed into
the sum $M=M^{(0)}+M^{(2)}+M^{(3)}+M^{(4)}$, such that  $\phi[M^{(n)}]=i^n M^{(n)}$.
\begin{exercise}
Find $M^{(n)}$ for $n=0,1,2,3$ explicitly in terms of $A,B,C,D,E$.
\end{exercise}
The invariant part $M^{(0)}$ is exactly $sp(2,2)\times sp(4)$. In particular we can decompose the current $J=J^{(0)}+J^{(1)}+J^{(2)}+J^{(3)}$ and define the action as
\beq\la{GS}
S=\frac{\sqrt\lambda}{4\pi}
\int {\rm str} \left(
J^{(2)}\wedge * J^{(2)}-J^{(1)}\wedge J^{(3)}\right)\;.
\eeq
\begin{exercise}
Show that $M^{(0)}\in sp(2,2)\times sp(4)$.
\end{exercise}
\begin{exercise}The fact that the action does not contain $J^{(0)}$ guarantees the local invariance of the action w.r.t. $sp(2,2)\times sp(4)$. Explain why.
\end{exercise}
The equations of motion which one can derive from the action \eq{GS} are
\beq\la{EOM}
\partial_\mu k_\mu=0\;\;,\;\;k_\mu=gK_\mu g^{-1}\;\;,\;\;K=J^{(2)}+\frac{1}{2}* J^{(1)}-\frac{1}{2}* J^{(3)}\;.
\eeq
One can also interpret $k_\mu$ as a Noether charge w.r.t to the global $PSU(2,2|4)$ symmetry $g\to h g$.
\begin{exercise}Derive $k_\mu$ from  Noether's theorem.
\end{exercise}

\section{Classical Integrability}
The equations of motion \eq{EOM} and the flatness condition:
\beq\la{flt}
dJ-J\wedge J=0
\eeq
can be packed into the 
flatness condition
of the 1-form
\beq\la{defA}
A(u)=J^{(0)}+\frac{u}{\sqrt{u^2-4g^2}} J^{(2)}-
\frac{2g}{\sqrt{u^2-4 g^2}} * J^{(2)}\;\;,\;\;u\in C
\eeq
where we use that classically we can set $J^{(1)}=J^{(3)}=0$, as these fermionic parts only become relevant at 1-loop level.
\newpage 
\begin{exercise}By expanding in Taylor series in $u$ show that each term in the expansion is zero as a consequence of \eq{EOM} and \eq{flt}, i.e.
\beq\la{flt2}
dA(u)-A(u)\wedge A(u)=0\;\;,\;\;\forall u.
\eeq
{Hint:} First verify \eq{flt2} for $u=0$.
For that you will have to project the equation \eq{flt} into $Z_4$ components first. For example
\beq\la{flt3}
dJ^{(0)}-J^{(0)}\wedge J^{(0)}-J^{(2)}\wedge J^{(2)}=0\;.
\eeq
\end{exercise}

The existence of the flat connection $A(u)$, depending on a spectral parameter $u$ implies integrability of the model
at least at the classical level. Note that \eq{flt2}\footnote{which in more familiar notations  becomes $F_{\mu\nu}=\partial_\mu A_\nu-\partial_\nu A_\mu +[A_\mu,A_\nu]=0$.} implies that $A(u)$ is a ``pure gauge" i.e. there exists a matrix valued function $G(\sigma,\tau,u)$ such that
\beq
A_\mu(u)=-G^{-1}\partial_\mu G\;.
\eeq
A way to build $G$ is to compute the Wilson line from some fixed point to $(\sigma,\tau)$
\beq
G(\sigma,\tau,u)={\rm Pexp}\int^{(\sigma,\tau)} A(u)\;.
\eeq
Using $G$ we can build
the monodromy matrix (which is a super matrix $(4+4)\times(4+4)$)
\beq
\Omega(u,\tau)=G^{-1}(0,\tau,u)G(2\pi,\tau,u)={\rm Pexp}\oint_\gamma A(u)\;.
\eeq
where $\gamma$ is a closed path starting and ending at some point
on the worldsheet and wrapping around once.
The flatness condition allows us to deform the contour freely 
provided the endpoints are fixed. 
Shifting the whole path in time will produce a similarity
transformation of $\Omega(u,\tau)$.
\begin{exercise}
Show that the eigenvalues of $\Omega(u,\tau)$ do not depend on $\tau$
if $A$ is flat. 
\end{exercise}
We denote the eigenvalues of $\Omega(u,\tau)$ as
\beq\la{qmomo}
\{
e^{i p_1},
e^{i p_2},
e^{i p_3},
e^{i p_4}|
e^{i p_{\hat 1}},
e^{i p_{\hat 2}},
e^{i p_{\hat 3}},
e^{i p_{\hat 4}}
\}\;.
\eeq
These functions of the spectral parameter $u$ are
called quasi-momenta. Since they do not depend on time they represent a generating function for conserved quantities. One can, for instance,
expand $p_i(u)$ in the Taylor series at large $u$ to obtain inifinitely many integrals of motion which leads to integrability of string theory. Below we study the analytic properties of the quasimomenta.

\paragraph*{``Zhukovsky" square roots}
All the quasimomenta have a square root singularity with the branch points at
$\pm 2g$ (inherited from the definition of $A$ \eq{defA}).
Note that the analytic continuation under the cut changes the sign of the terms
with $J^{(2)}$ in \eq{defA} which is in fact equivalent to applying the $Z_4$ automorphism.

At the same time one can show that
\beq\la{COC}
C^{-1}\Omega(u)C=\tilde\Omega^{-ST}(u)\;\;,\;\;C=
\left(
\bea{c|c}
E&0\\ \hline
0& E
\eea
\right)
\eeq
where $\tilde\Omega(u)$ denotes the analytic continuation of
$\Omega(u)$ under the cut $[-2g,2g]$.
\begin{exercise}
Show that 
\beq
C^{-1} M C=- \phi[M]^{ST}
\eeq
where $ST$ denotes super transpose is defined as
$
\left(
\bea{c|c}
A&B\\ \hline
C&D
\eea
\right)^{ST}\equiv \left(
\bea{c|c}
A^T&C^T\\ \hline
-B^T&D^T
\eea
\right)
$.
Use this to show that $C^{-1}AC=-\tilde A^{ST}$ (where tilde denotes analytic continuation under the branch cut $[-2g,2g]$). Then prove \eq{COC}. 
\end{exercise}
Equation \eq{COC} implies that the eigenvalues of $\Omega(u)$ are related to the eigenvalues of $\tilde\Omega(u)$ by inversion and possible permutation. This statement in terms of the quasimomenta \eq{qmomo} tells us that the analytic continuation of the quasimomenta i.e. $\tilde p_a(u)$ and $\tilde p_{\hat i}(u)$
results in the change of sign and possible reshuffling.
The exact way they reshuffle can be determined by considering some
particular classical solutions and building the quasimomenta explicitly.
Some examples can be found in \cite{Gromov:2007aq}. Since all the classical solutions 
are related to each other continuously one finds that\footnote{It is also possible to shift the quasimomenta by $2\pi m$ where $m$ is integer. This is indeed the case for $p_i$ for the classical solutions which wind in $S^5$ and $m$ gives their winding number. The $AdS^5$ quasimomenta still satisfy \eq{perm}.}
\beqa\la{perm}
\tilde p_{\hat 1}(u)=-p_{\hat 2}(u)\;\;,\;\;
\tilde p_{\hat 2}(u)=-p_{\hat 1}(u)\;\;,\;\;
\tilde p_{\hat 3}(u)=-p_{\hat 4}(u)\;\;,\;\;
\tilde p_{\hat 4}(u)=-p_{\hat 3}(u)\;.
\eeqa
This property will play a crucial role in the QSC construction as we discuss in the next section. One can consider \eq{perm} as a manifestation of $Z_4$ symmetry of the action.

\paragraph*{Large $u$ asymptotics and quantum numbers}
Another important property of the quasimomenta is that the quantum numbers of the state can be read off from their values at infinity.
To see this notice the following property
\beq
A=-g^{-1}\left(d+*k\frac{2g}{u}\right)g
\eeq
where $k_\mu$ is the Noether current defined in \eq{EOM}. 
This implies that 
\beq
\Omega=-g^{-1}\left(1+\frac{2g}{u}\int_0^{2\pi}d\sigma k_\tau\right)g
\eeq
using that the charge $Q_{\rm Noether}=2g\int_0^{2\pi}k_\tau d\sigma$ we immediately get
\beq
\left(\bea{c}
 p_{\hat 1}\\
 p_{\hat 2}\\
 p_{\hat 3}\\
 p_{\hat 4}\\  \hline
 p_1\\
 p_2\\
 p_3\\
 p_4\\
\eea\right) \simeq
\frac{1}{2 u}
\left(\bea{l}
+\Delta-S_1+S_2 \\
+\Delta+S_1 -S_2 \\
-\Delta-S_1  -S_2 \\
-\Delta+S_1 +S_2 \\  \hline
+J_1+J_2-J_3  \\
+J_1-J_2+J_3 \\
-J_1+J_2 +J_3 \\
-J_1-J_2-J_3
\eea\right)\,. \label{inf}
\eeq
where the r.h.s. comes from the diagonalization of $Q_{\rm Noether}/u$ (in the fundamental representation). Here $J_i$ are integer R-charges (which map to the scalar fields in gauge theory), $S_1,S_2$ are integer Lorentz charges (corresponding to the covariant derivatives)
and $\Delta$ is the dimension of the state, i.e. its energy. 
Again we will see the quantum counterpart of this formula when we discuss QSC construction in the next section.
\paragraph*{Action variables and WKB quantization}
Another reason the quasimomenta were introduced is because they allow us to define the action variables very easily. For non-trivial solutions the quasimomenta have additional quadratic branch cuts, which come from the diagonalization procedure. The integrals around these cuts give the action variables~\cite{Dorey:2006zj}\footnote{this property fixes the choice of the spectral parameter $u$, which otherwise can be replaced by any $f(u)$.}
\beq\la{Actionv}
I_{ C}=\frac{1}{2\pi i}\oint_C p_A(u) du\;,
\eeq
where $C$ is some branch cut of $p_A(u)$.
Here $A$ can take any of $8$ values.
In the Bohr-Sommerfeld quasi-classical quantization procedure one simply imposes $I_C\in Z$ to get the first quantum correction. For example in \cite{Gromov:2007aq} this property was used to obtain the 1-loop quantum spectrum of the string. 

\section{Quasimomenta and the Strong Coupling Limit of QSC}
To understand how the quansimomenta we introduced above are related to the $Q$-functions
from the previous section we are going to first get an insight from the harmonic oscillator.
Reconstructing the $\psi$ from $p$, but inverting the relation \eq{pdef} we get
\beq\la{psix}
\psi(x)=e^{-\frac{m\omega x^2}{2\hbar}}Q(x)=e^{\frac{i}{\hbar}\int^x p(x) dx}\;.
\eeq
Similarly to what we found in \eq{Actionv} we also had
\beq\la{HOW}
N=\frac{1}{2\pi i}\frac{i}{\hbar}\oint_C p(x)dx\;,
\eeq
which allows us to identify $\frac{i x}{\hbar}\to u$
so that \eq{HOW} and \eq{Actionv} become really identical.
Under this identification we can deduce from \eq{psix}
\beq
Q_A\simeq\exp\left(\int^u p_A(v) dv\right)\;.
\eeq
This naive argument indeed produces the right identification for the strong coupling 
limit (i.e. $g\to\infty$)
of
 ${\bf P}_a$ and ${\bf Q}_i$ functions introduced earlier. More precisely we get:
\beqa\la{expP}
{\bf P}_a\sim\exp\left(-\int^u p_a(v) dv\right)\;\;&,&\;\;
{\bf P}^a\sim\exp\left(+\int^u p_a(v) dv\right)\\
{\bf Q}_i\sim\exp\left(-\int^u p_{\hat i}(v) dv\right)\;\;&,&\;\;
{\bf Q}^i\sim\exp\left(+\int^u p_{\hat i}(v) dv\right)\;.
\eeqa
Note that at the leading classical level we do not control the preexponential factors and they may contain some order $1$ powers of $u$.
From that we can immediately draw a number of important consequences:
\begin{itemize}
\item We can deduce that large $u$ asymptotics of $\bP_a$ and $\bQ_i$ from \eq{inf} are
of the form $u^{Q_{\rm Noether}/2}$.
\item We can no longer expect that $\bP_a$ or $\bQ_i$ are polynomials as the expressions
\eq{expP} have Zhukovsky branch cuts $[-2g,2g]$.
\item From \eq{perm} we can deduce the following analytic  continuation under the branch cut
\beq
\tilde \bQ_i \sim\exp\left(+\int^u \tilde p_{\hat i}(v) dv\right)=
\exp\left(-\int^u p_{\hat\phi_i}(v) dv\right)\sim \bQ^{\phi_i}
\eeq 
where $\phi_i$ is determined by \eq{perm} to be $\phi_1=2,\;\phi_2=1,\;\phi_3=4,\;\phi_4=3$. So more explicitly we should have the following monodromies
\beq\la{classical}
\tilde \bQ_1=\bQ^2\;\;,\;\;\tilde \bQ_2=\bQ^1\;\;,\;\;\tilde \bQ_3=\bQ^4\;\;,\;\;\tilde \bQ_4=\bQ^3\;.
\eeq
These relations remain almost intact at the quantum level. The only improvement one should make is to complex conjugate the r.h.s., as at the quantum level the $Q_i$ are not real
\beq\la{gluing}
\tilde \bQ_1=\bar \bQ^2\;\;,\;\;\tilde \bQ_2=\bar \bQ^1\;\;,\;\;\tilde \bQ_3=\bar \bQ^4\;\;,\;\;\tilde \bQ_4=\bar \bQ^3\;.
\eeq
The reason for the complex conjugation will become clear in the next Chapter.
\end{itemize}
To conclude this section we notice that we managed to get all the crucial additional information we have to add to the QQ-relations from just  classical limit. Namely, the existence of the Zhukovsky cut and the ``gluing" conditions \eq{gluing}. In the next chapter we combine all the information together and give the complete description of the spectrum of $N=4$ SYM by means of the QSC.

\chapter{QSC Formulation}\la{ch:4}
The goal of this section is to summarize the insights we got from the classical limit and from the spin chains and to motivate further the analytic properties of the basic ${\bf P}_a,\;{\bf P}^a,\;{\bf Q}_i,\;{\bf Q}^i$ Q-functions.

\section{Main QQ-Relations}
The Q-functions of the ${\mathcal N}=4$ SYM satisfy exactly the same QQ-relations as those of the $SU(4|4)$ spin chain. So we simply summarize the most important relations from Sec.\ref{sec:PSU224QQ} here to make this section self-contained
\beqa
\la{allQQ1}&&Q_{a| i}^+-Q_{a| i}^-={\bf P}_a{\bf Q}_i\;\;,\\
&&{\bf P}_a{\bf P}^a=\bQ_i \bQ^i=0\;,\\
\la{allQQ2}&&{\bf Q}_i=-{\bf P}^a Q_{a|i}^+
\;\;,\\
\la{allQQ22}
&&{\bf Q}^i=+
{\bf P}_a Q^{a|i+}\;\;,\\
\la{allQQ3}&&Q^{a|i}=-(Q_{a|i})^{-t}\;\;.
\eeqa
We also note that the first identity \eq{allQQ1} can be combined with \eq{allQQ2}
into
\beq
\la{allQQ0}Q_{a| i}^+-Q_{a| i}^-=-{\bf P}_a{\bf P}^b{Q}^+_{b|i}\;.
\eeq
This relation tells us that we can use $8$ functions $\bP_a$ and $\bP^a$
as the basis to reconstruct all other Q-functions i.e. we can in principle solve \eq{allQQ0} 
in terms of  $\bP$'s (we will have an example in the next sections). Then we can use $Q_{a|i}$ to find $Q^{a|i}$ as its inverse \eq{allQQ3}. Then one can reconstruct $\bQ_i$ and
$\bQ^i$ using \eq{allQQ2} and \eq{allQQ22}. 

The advantage of this choice of basis is, as we explain below, due to the fact that the analytic properties of $\bP_a$ and $\bP^a$ are the simplest among all $Q$-functions and they can be very efficiently parameterized.
\section{Large $u$ Asymptotic and the Quantum Numbers of the State}
The large $u$ asymptotics of $\bP$'s and $\bQ$'s can be deduced from their classical limit
\eq{expP} and \eq{inf}. The main complication here is that in the non-twisted theory there are some additional powers of $u$ comming from the pre-exponent of \eq{expP}, which modify the asympotic by $\pm 1$. To fix the asymptotic completely one can make comparison with the Asymptotic Bethe Ansatz of Beisert-Staudacher, which can be derived as a limit of QSC.
We don't discuss this calculation here but this was done in detail in the original paper \cite{Gromov:2014caa}.
Here we just quote the result
\begin{eqnarray}\label{largeu2}
\bP_a\simeq A_a\, u^{-\tilde M_a}\,,\ \ \bQ_{i}\simeq B_i\,u^{\hat M_i-1}\,,\qquad   \bP^a\simeq A^a\, u^{\tilde M_a-1}\,,\ \ \bQ^{i}\simeq B^i\,u^{-\hat M_i}\,,
\end{eqnarray}
where
\begin{align}
\label{relMta}
\tilde M_a=&\left\{\frac{J_1+J_2-J_3+2}{2}
    ,\frac{J_1-J_2+J_3}{2}
    ,\frac{-J_1+J_2+J_3+2}{2}
    ,\frac{-J_1-J_2-J_3}{2}
    \right\}\\
\hat M_i=&\left\{\frac{\Delta -S_1-S_2+2}{2},
\frac{\Delta +S_1+S_2}{2}
   ,\frac{-\Delta
   -S_1+S_2+2}{2}  ,\frac{-\Delta
   +S_1-S_2}{2}  \right\}
\label{M-ass}\end{align}
we see that indeed the asymptotics are consistent with what we found in the classical limit.
Another way to understand the shift by $\pm 1$ in the asymptotic is to consider a more general twisted theory. The twists (like the parameter $\phi$ we introduced in the spin chain section) remove many degeneracies\footnote{see \cite{Kazakov:2015efa} for more details about the twisted version of QSC.}. For example without the twist
the leading asymptotic in the l.h.s. of \eq{allQQ1} cancels and one needs to know the subleading term to deduce the asymptotic of the r.h.s. This does not happen in the twisted case when $Q_{a|i}\sim e^{\phi_{a,i}u}u^{M_{a,i}}$ and the asymptotic behaves more predictably. As a result in the twisted theory there are no $\pm 1$ shifts w.r.t. the classical limit asymptotic and one can alternatively derive \eq{largeu2} by considering first the twisted ${\mathcal N}=4$ SYM and then removing them.

\paragraph*{Finding normalization of $\bP$ and $\bQ$}
We will see in the next section that in the near BPS limit $\bP$ and $\bQ$ become small
which will allow us to solve the QSC exactly at finite coupling. In order to see this we derive a more general result for the coefficients $A_a,\;A^a$ and $B_i,\;B^i$ from \eq{largeu2}.
\begin{exercise}
Use \eq{allQQ1} and \eq{largeu2} to show that
\begin{eqnarray}\label{Qkjlargeu}
Q_{a|j}\simeq
-i\, A_a\,B_j\,\frac{u^{-\tilde M_a+\hat M_j}}{-\tilde M_a+\hat M_j}\,.
\end{eqnarray}
\end{exercise}

From  \eq{Qkjlargeu} we can fix the constants \(A^a\) and \(B^i\)
in terms of \(\tilde M_a\) and \(\hat M_j\). Substituting the asymptotic 
\eq{Qkjlargeu} into \eq{allQQ2} we get
\beq
-A^au^{\tilde M_a-1}\left(-i\, A_a\,B_j\,\frac{u^{-\tilde M_a+\hat M_j}}{-\tilde M_a+\hat M_j}\right)=
B_j u^{\hat M_j-1}\;,
\eeq
which simplifies to
\beq
-1= i\,\sum_{a=1}^4\frac{A^aA_a}{\tilde M_a-\hat M_j}\;,
\eeq
which allows us to find the combinations $A^1A_1,\;A^2A_2$
and so on.
Solving this linear system we find
\begin{eqnarray}\label{AABB0}
A^{a_0}A_{a_0}=i\frac{\prod\limits_{j}(\tilde M_{a_0}-\hat M_j)}{\prod\limits_{b\neq a_0}(\tilde M_{a_0}-\tilde M_b)}\,,\;\;
                B^{j_0}B_{j_0}=i\frac{\prod\limits_{a}(\hat M_{j_0}-\tilde M_a)}{\prod\limits_{k\neq
{j_0}}(\hat M_{j_0}-\hat M_k)}\;,\;\; a_0,j_0=1\dots4
\end{eqnarray}
(with no summation over \(a_0\) or \(j_0\) in l.h.s.).
\begin{exercise}
Derive the relations \eq{AABB0}.
\end{exercise}
Interestingly the condition that all $A^{a_0}A_{a_0}=0$ 
for all $a_0$ singles out the BPS states
with protected dimension (which works for physical and even
non-physical operators like in the BFKL regime).
\begin{exercise}
Find all solutions of $A^{a_0}A_{a_0}=0,\;\forall a_0$ in terms of $J_i,\;S_i$ and $\Delta$.
\end{exercise}
Next we investigate the cut structure of $\bP_a$ and $\bQ_i$.

\section{Analytic Structure of Q-functions}
In this section we deduce the analytic properties of $\bP_a$ and $\bQ_i$
functions following a maximal simplicity principle, i.e. we assume simplest possible 
analytical properties which do not contradict the classical limit and the structure of the QQ-system.
In Sec.\ref{sec:classics} from the 
 strong coupling analysis we deduced that $\bP_a$
and $\bQ_i$ should have cuts with branch points
at $\pm 2g$ (to recall $g=\frac{\sqrt{\lambda}}{4\pi}$ where $\lambda$ is the 't Hooft coupling). 
We can assume that $\bP_a$ should
have just one single cut $[-2g,2g]$. Note that since $\bP^a$
is related to $\bP_a$ by the symmetry of flipping the Dynkin 
diagram upside-down it should also have the same analytic properties.

Note that $\bQ_i$ (and $\bQ^i$) cannot have the same analytic properties as $\bP$'s. Indeed, in general $\Delta$ in the asymptotic of $\bQ_i$ is not-integer and thus we must have a nontrivial monodromy around infinity.\footnote{Depending on the values of $J_i$ there could be a similar issue with $\bP_a$ as the asymptotic could contain half-integer numbers. Strictly speaking $\bP_a$ could have an extra cut going to infinity which would disappear in any bi-linear combinations of $\bP$s.} 
The simplest way to gain such a monodromy is to choose the branch cut to close through infinity i.e. we can assume that $\bQ_i$ and $\bQ^i$
have
a ``long" branch-cut $(-\infty,-2g]\cup[+2g,+\infty)$.
This simple argument leads us to the simple analyticity picture Fig.\ref{cuts}, which historically was derived using the TBA approach in \cite{Gromov:2014caa}. 

\begin{figure}
\begin{center}
\def\svgwidth{\textwidth}
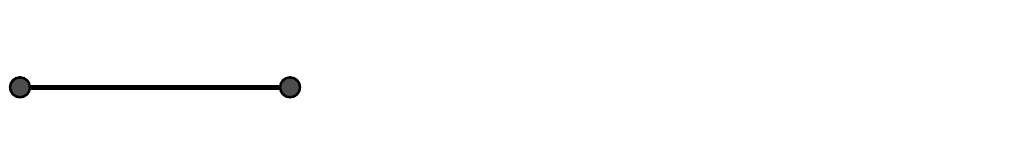
\end{center}
\caption{\la{cuts}cut structure of $\bP_a,\;\bP^a$ and $\bQ_i,\;\bQ^i$.}
\end{figure}

Note that $\bP$ and $\bQ$ are additionally constrained to be a  part of the same Q-system. This makes it very inconvenient to have different conventions for the choice of the branch cuts so we may need to look under the long cut of $\bQ$.
A simple way to explore the space under the cut of $\bQ$ is to use the QQ-relation \eq{allQQ0} written in the form
\beq
Q_{a|i}^++\bP_a\bP^b Q_{b|i}^+=Q_{a|i}^-
\eeq
which implies that
\beqa\la{Qi}
\nn Q_i&=&-\bP^aQ^+_{a|i}=-\bP^a(\delta_a^b+\bP^{[+2]}_a \bP^{b[+2]})Q_{b|i}^{[3]}\\
&=&-\bP^a(\delta_a^b+\bP^{[+2]}_a \bP^{b[+2]})
(\delta_b^c+\bP^{[+4]}_b \bP^{c[+4]})Q_{c|i}^{[5]}=\dots\;.
\eeqa
First we note that from the formal solution \eq{Qaiformal} we can always assume that $Q_{a|i}$ is regular in the upper half plane. From that we see that \eq{Qi} implies that $\bQ_i$ has an infinite ladder of cuts. 
The first term in the last line of \eq{Qi} has a cut 
at $[-2g,2g]$,
the second has the cut at $[-2g-i,2g-i]$ and so on. See Fig.\ref{Qishort}. 
The puzzle is how to make this structure of cuts compatible
 with the initial guess that $\bQ_i$
has only one cut going to infinity. In fact there is no contradiction so far as in order to see the infinite ladder of the cuts we should go 
to the right from the branch point at $2g$ i.e. under the long cut.
At the same time if we want to go to the lower half plane avoiding the long cut we should go under the first short cut. What is expected to be seen under the first short cut is no branch point singularities below the real axis. Thus if we denote by $\tilde \bQ_i$ the analytic continuation of $\bQ_i$
under the first cut it will have no branch cut singularities below the real axis (see Fig.\ref{Qishort}).

\begin{figure}[ht]
\begin{center}
\def\svgwidth{0.7\textwidth}
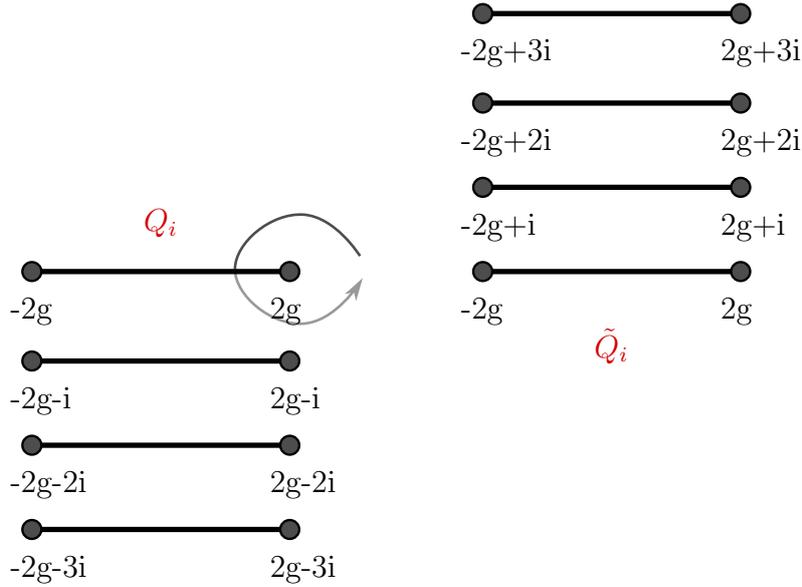
\end{center}
\caption{Analytic structure of $Q_i$ under its long cut.\la{Qishort}}
\end{figure}

From Fig.\ref{Qishort} we notice an obvious asymmetry of the upper half plane and the lower half plane. Indeed the function $\bQ_i$, which is a part of Q-system, is analytic in the upper half plane, whereas $\tilde \bQ_i$, which does not necessarily satisfy any QQ-relation is analytic in the lower half plane. I.e. building the Q-system we decided to keep all Q-functions analytic above  the real axis and now we potentially lost the symmetry under complex conjugation which can be linked to unitarity of the theory. To reconstruct the symmetry we have to impose the ``gluing conditions".

\section{Gluing Conditions}
In this section we address an imperfection of our construction where the upper half plane plays a more important role from a QQ-relations point of view. To exchange the upper and lower halves we can complex conjugate Q-functions. This procedure does not affect $\bP_a$ and $\bP^a$ much,
depending on the normalization constant they can at most change their signs. For $\bQ_i$ the complex conjugation seems to be more dramatic as the ladder of branch cuts going down will now go up. Simply multiplying $\bQ_i$ by a constant would not undo the complex conjugation, however, if we also analytically continue $\bar \bQ_i$ under  the  first branch cut we actually get a very similar analytic structure to the initial  $\bQ_i$!
I.e. the
complex conjugation and the analytic continuation should give us back either some $\bQ_i$ or $\bQ^i$. To determine which of $\bQ$ could do the job we recall that in the classical limit we obtained \eq{gluing} and in accordance with that we impose the following gluing conditions\footnote{for physical operators $\bQ_2$ and $\bQ_4$ 
can mix with $\bQ_1$ and $\bQ_3$ as they are growing faster
and have the same non-integer part in the asymptotic (similarly $\bQ^3$ and $\bQ^1$ are defined modulo $\bQ^2$ and $\bQ^4$). As a result
they are ambiguously defined. Fortunately, we don't have to impose all the $4$ of the
 gluing conditions and it is sufficient to
use another pair of equations.}
\beq\la{gluing2}
\boxed{\tilde \bQ_1\propto\bar \bQ^2\;\;,\;\;\tilde \bQ_2\propto\bar \bQ^1\;\;,\;\;\tilde \bQ_3\propto\bar \bQ^4\;\;,\;\;\tilde \bQ_4\propto\bar \bQ^3\;.}
\eeq
Together with \eq{allQQ1}-\eq{allQQ3}, \eq{gluing2} constitutes
the closed system of QSC equations. It is rather nontrivial that these equations only have a discrete set of solutions (and so far there is no mathematically rigorous proof of this). To demonstrate this we consider some simple examples in the next section and also implement an algorithm which allows us to find solutions numerically.


\section{Left-Right Symmetric Sub-Sector}
In many situations it is sufficient to restrict ourselves to a subset of all states which has an additional symmetry. The left-right (LR) symmetric sub-sector, which includes $su(2)$ and $sl(2)$ sub-sectors, contains the states preserving the upside-down symmetry of the Dynkin diagram, i.e. the states which should have $J_3=0,\;S_2=0$.
To understand what we should expect in this case consider the bosonic subgroup $SO(4,2)\times SO(6)$. The $SO(6)$ Dynkin diagram has $3$ nodes and imposing the LR symmetry would imply that the nodes $1$ and $3$ are indistinguishable which reduces the symmetry to $SO(5)$ (see Fig.\ref{so5}). 
\begin{figure}
\begin{center}
\def\svgwidth{0.6\textwidth}
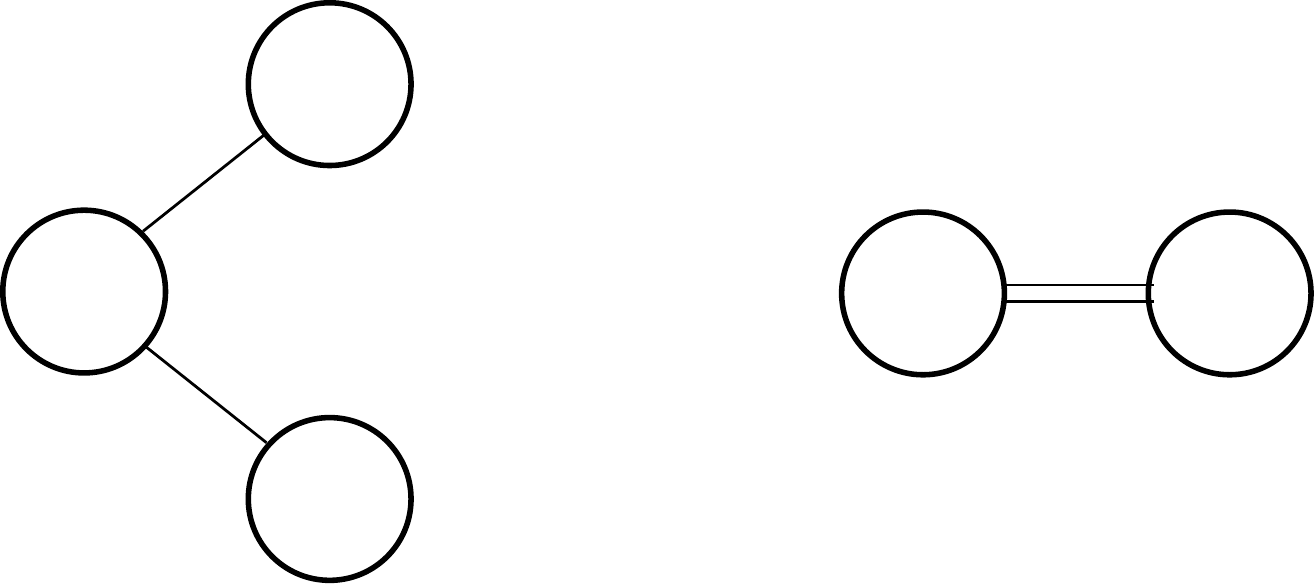
\end{center}
\caption{\la{so5} Under identification of the upper and lower nodes the $SO(6)$ Dynkin diagram (on the left) becomes the $SO(5)$ Dynkin diagram (on the right).}
\end{figure}
In order to break $SO(6)$ to $SO(5)$ it is sufficient to select some preferable direction in the vector $6$D representation. Our $\bP_a$
and $\bP^a$ are in $4$D fundamental and anti-fundamental representations of $SO(6)$. The vector representation can be realized as anti-symmetric tensors with two fundamental indexes $A_{ab}$
and so we can pick a direction to break $SO(6)$ to $SO(5)$ by picking a particular anti-symmetric tensor $\chi_{ab}$ which can be used to relate fundamental and anti-fundamental representations, i.e. can be used to lower the indexes. In this sub-sector we will get
\beq\la{Plower}
\bP_a=\chi_{ab}\bP^b\;.
\eeq
Since we have already selected the order of $\bP$'s by assigning their asymptotic we can see that the only non-zero components of $\chi$,
consistent  with the asymptotic of $\bP$ are $14,23,32,41$. Finally we still have freedom to rescale $\bP_a$ to bring $\chi_{ij}$ to the conventional form 
\beq
\chi_{ab}=\left(
\bea{cccc}
0&0&0&1\\
0&0&-1&0\\
0&1&0&0\\
-1&0&0&0
\eea
\right)\;.
\eeq
By the same argument we should impose
\beq\la{Qlower}
\bQ_i=\chi_{ij}\bQ^j\;
\eeq
for the same tensor $\chi_{ij}$. 

\chapter{QSC - analytic examples}\la{ch:5}
In this section we consider an example where the QSC can be solved analytically at finite coupling. It is unfortunate that the analytical solutions for physical operators are rather complicated. It is possible to get the solution perturbatively at weak coupling, but this already  involves computer algebra. Here instead we consider a non-local operator which can be understood as an analytic continuation of twist-$J$ states.
The twist operators are the states with $J_1=J$, $J_2=J_3=0$ and $S_1=S,\;S_2=0$. They belong to the LR symmetric subsector described in the previous section and below we give the description of the $sl(2)$ sector to which these states also belong in the next section.

\section{$sl(2)$ Sector}
We discuss the simplifications which arise in the $sl(2)$ sector.
As the $sl(2)$ sector is inside the LR subsector we can restrict ourselves to the Q-functions with lower indexes due to \eq{Qlower} and \eq{Plower}.
The asymptotics of $\bP_a$  \eq{largeu2} become 
\beq\la{Pasm}
\bP_1=A_1 u^{-L/2-1}\;\;,\;\;
\bP_2=A_2 u^{-L/2}\;\;,\;\;
\bP_3=A_3 u^{+L/2-1}\;\;,\;\;
\bP_4=A_4 u^{+L/2}\;.
\eeq
Similarly for $\bQ_i$
\beqa
\nn&&\bQ_1=B_1 u^{+(\Delta-S)/2}\;\;,\;\;
\bQ_2=B_2 u^{+(\Delta+S)/2-1}\;\;,\\
&&\bQ_3=B_3 u^{-(\Delta+S)/2}\;\;,\;\;
\bQ_4=B_4 u^{-(\Delta-S)/2-1}\;\;.
\eeqa
Also we write \eq{AABB0} explicitly for this case
\beqa\la{AAsl2}
\nn A_1A_4&=&-\frac{i (-\Delta +L-S+2) (-\Delta +L+S) (\Delta +L-S+2) (\Delta +L+S)}{16 L (L+1)}\\
A_2A_3&=&-\frac{i (-\Delta +L-S) (-\Delta +L+S-2) (\Delta +L-S) (\Delta +L+S-2)}{16 (L-1) L}
\eeqa
and
\beqa\la{BBsl2}
\nn B_1B_4&=&
-\frac{i (-\Delta +L+S-2) (-\Delta +L+S) (\Delta +L-S) (\Delta +L-S+2)}{16 \Delta
   (S-1) (-\Delta +S-1)}\\
B_2B_3&=&
+\frac{i (\Delta -L+S-2) (\Delta -L+S) (\Delta +L+S-2) (\Delta +L+S)}{16 \Delta
   (S-1) (\Delta +S-1)}\;.
\eeqa
We can see that both $A_a A^a$ and $B_aB^a$ vanish for $\Delta\to L,\;S\to 0$. The reason for this is that $S=0$ is the BPS protected state and the vanishing of the coefficients indicates the shortening of the multiplet. At the same time when $\bP$ and $\bQ$ are small we get an enormous simplification as we show in the next section where we consider a near BPS limit where $S$ is small. 
 
\section{Analytic Continuation in $S$}\la{sec:cont}
In this section we introduce an analytic continuation in the Lorentz spin $S_1=S$, which for local operators must be integer. The analytic continuation in the spin plays an important role as it links BFKL and DGLAP regimes or high energy scattering in QCD\footnote{for the applications of QSC in this regime see \cite{Alfimov:2014bwa,Gromov:2015vua}}. We leave the questions related to the physics of hight energy scattering outside these lectures and describe in detail the analytic continuation in $S$ from QSC point of view.

The simplest way to describe the analytic continuation is by considering the gluing conditions \eq{gluing2}, which for LR-symmetric sector reduce to just two
\beq\la{gluing}
\tilde \bQ_1\propto\bar \bQ_3\;\;,\;\;
\tilde \bQ_2\propto\bar \bQ_4
\;,
\eeq
since the two others gluing conditions follow by taking complex conjugate and analytically continue the above two conditions.

Also, as we will see that from the numerical analysis in the next section, these two conditions are not independent and only one of them is sufficient to build the spectrum. 
At the same time imposing both conditions \eq{gluing} leads to the quantization of the charge $S_1$ whereas keeping only the first condition
$\tilde \bQ_1\propto\bar \bQ_3$ allows us to have $S_1$ non-integer\footnote{one can show that the second condition necessary leads to the quantization of $S$~\cite{GA}. It could be simpler to check this numerically with the code we explain in the next Chapter.}! However,
this will modify the second gluing condition. To constrain the possible form of the modified gluing conditions we denote
\beqa\la{Mdef}
\tilde \bQ_i(u)&=& {M_{i}}^j(u)\bar \bQ_j(u)\;\;,\;\;
{M_i}^j(u)=
\left(\bea{cccc}
0&0&{M_{1}}^3&0\\
{M_{2}}^1&{M_{2}}^2&{M_{2}}^3&{M_{2}}^4\\
{M_{3}}^1&0&0&0\\
{M_{4}}^1&{M_{4}}^2&{M_{4}}^3&{M_{4}}^4
\eea\right)_{ij}\;.
\eeqa
Since the gluing condition tells us that $\tilde \bQ_i$ is essentially the same as $\bQ^i$
up to a possible symmetry of Q-system transformation we should assume that ${M_i}^j(u)$ is an i-periodic function of $u$: ${M_i}^j(u+i)={M_i}^j(u)$. Furthermore,
since ${M_i}^j(u)$ relates two functions which are both analytic in the lower half plane it should be analytic.
\newpage 
\begin{exercise}
Use periodicity of ${M_i}^j$ and equation \eq{Mdef} to find $M_k^j$
explicitly in terms of $\tilde \bQ_k(u),\tilde \bQ_k(u+i),\tilde \bQ_k(u+2i),\tilde \bQ_k(u+3i)$ and $\bar \bQ_j(u),\bar \bQ_j(u+i),\bar \bQ_i(u+2i),\bar \bQ_j(u+3i)$. From that relation you can see that ${M_{k}}^j$ does not have any branch-cuts, but could possibly have poles. However, existence of poles would contradict the power-like asymptotic of $\bar \bQ_j(u)$ and analyticity of $\tilde \bQ_i(u)$ as we will have to conclude that $\bar \bQ_j$ has infinitely many zeros in the  lower half plane, which is impossible with a power-like asymptotic. 
\end{exercise} 
Armed with the new knowledge of regularity of $M$ we can analytically continue both sides of \eq{Mdef} and complex conjugate them to find the following condition on the matrix $M$: 
\beqa\la{Mreality}
\bar M(u)=M^{-1}(u)\;.
\eeqa
Another constraint comes from the LR-symmetry of the state, which tells us that $\bQ_i=\chi_{ij}\bQ^j$, where $\bQ^j$ is a tri-linear combination of $\bQ_i$ as in \eq{QQ3}. So using that we get from \eq{Mdef}
\beq\la{Mdef2}
\tilde \bQ^l(u)=(\chi^{-1})^{li} {M_{i}}^j\chi_{jk}\bar \bQ^k(u)\;\;,\;\;
\eeq
at the same time we can use \eq{QQ3} and \eq{hdu} to rewrite the r.h.s. as a combination of $3$ $\tilde \bQ_i$ and then apply the initial \eq{Mdef}; this results in the following equation
\beq\la{Mdef3}
\tilde \bQ^i(u)= -{\rm det}(M){(M^{-1})_{j}}^i\bar \bQ^j(u)\;.
\eeq
Comparing \eq{Mdef2} and \eq{Mdef3} we get 
\beq\la{Mconstr}
(\chi^{-1})^{li} {M_{i}}^j\chi_{jk}{M_n}^{k}=-\det(M)\delta^{l}_n\;,
\eeq
\begin{exercise}
Derive \eq{Mconstr} by combining \eq{Mdef2} and \eq{Mdef3}.
\end{exercise}
or in matrix form
\beq
M\chi M^T=-\chi\;{\rm det}(M)\;. 
\eeq
Which implies in particular that $\det(M)=\pm 1$.
Imposing \eq{Mconstr} and \eq{Mreality}
we obtain that $M$ should reduce to the following form
\beqa
{M_i}^j(u)=
\left(\bea{crrr}
0&0&\alpha&0\\
\beta&0&\gamma&-\bar{\alpha}\\
\frac{1}{\bar{\alpha}}&0&0&0\\
\frac{\gamma}{\alpha\bar\alpha}&-\frac{1}{\alpha}&\bar{\beta}&0
\eea\right)_{ij}\;,
\eeqa
with real $\gamma$, which results in the following two independent gluing conditions
\beqa\nn\la{gluing}
\tilde\bQ_1&=&\alpha\bar \bQ_3\;,\\ \la{gluingF}
\tilde\bQ_2&=&\beta\bar \bQ_1+\gamma \bar\bQ_3-\bar{\alpha}\bar\bQ_4\;.
\eeqa
Since $\alpha$ appears both in numerator and denominator it cannot be a non-trivial function of $u$ as it would create poles.
At the same time $\beta$ and $\gamma$ can be nontrivial periodic functions of $u$. For the case of the twist two operators ${\rm tr}Z D_-^S Z$\footnote{$Z$ is a complex scalar of the theory, $D_-$ is a light-cone covariant derivative.} with non-integer Lorentz spin $S$ we will verify numerically that $\gamma$ is a constant
and $\beta=\beta_1+\beta_2\cosh(2\pi u)+\beta_3\sinh(2\pi u)$. For integer $S$ both $\gamma$ and $\beta$ vanish.
Using this gluing matrix one can compute the BFKL pomeron/odderon eigenvalue by analytically continuing to $S\sim-1$.

\section{Slope Function}\la{slope}
Having the possibility of having the Lorentz spin $S$ non-integer
allows us to study the near-BPS regime $S\to  0$ analytically.
In this section we will compute the first term linear in $S$ called to slope function~\cite{Basso:2011rs} analytically to all orders in $g$.
This calculation was precented originally in~\cite{Gromov:2014bva}
in a slightly different form, there also the next term in small $S$ expansion was derived. Here we adopt more widely accepted notation of~\cite{Gromov:2014caa}, which are different from~\cite{Gromov:2014bva}.

The main simplification in this limit is due to the scaling of $\bP_a$ and $\bQ_i$ with $S\to 0$ and $\Delta=L+e S$ where $e\sim 1$, which can be deduced from the scaling of $A_a$
and $B_i$ \eq{AAsl2} and \eq{BBsl2}:
\beqa\la{AAvals}
A_1A_4\simeq-B_1B_4\simeq-\frac{i}{2}(1-e)S\;\;,\;\;
A_2A_3\simeq-B_2B_3\simeq-\frac{i}{2}(1+e)S\;.
\eeqa
From that we can deduce that $\bP_a$ and $\bQ_i$ both scale as $\sqrt S$.
This assumption in the main simplification -- the equation for $Q_{a|i}$ \eq{allQQ1} becomes simply
\beq
Q_{a|i}^+-Q_{a|i}^-\simeq 0\;,
\eeq
i.e. $Q_{a|i}$ is a constant!
To find which constants they are we can simply use the general formula \eq{Qkjlargeu} which in our limit gives
\beq
Q_{a|j}=\left(
\begin{array}{cccc}
-\frac{2 i A_1 B_1}{(e-1) S} & 0
& 0 & 0 \\
0 & -\frac{2 i A_2 B_2}{(e+1) S}
& 0 & 0 \\
0 & 0 & \frac{2 i A_3 B_3}{(e+1)
S} & 0 \\
0 & 0 & 0 & \frac{2 i A_4
B_4}{(e-1) S} \\
\end{array}
\right)\;.
\eeq
Using the rescaling symmetry\footnote{\la{rescaling}We can rescale $\bP_1\to f\bP_1$ and $\bP_2\to g\bP_2$, rescaling simultaneously $\bP_3\to 1/f\bP_1$ and $\bP_4\to1/g\bP_4$ and similar for $\bQ_i$. In addition for $\bP$'s only we have the freedom $\bP_3\to\bP_3+\gamma_2\bP_2-\gamma_1\bP_1$
and $\bP_4\to\bP_4+\gamma_3\bP_1+\gamma_1\bP_2$ for some constants $\gamma_n$, this ambiguity is resolved in the twisted theory. These transformations are the most general which preserve $\chi_{ij}$ tensor and do not modify the asymptotic of $\bP$'s.} we can set $B_1=i A_4,\;B_2=i A_3,\;B_3=iA_2,\;B_4=iA_1$ giving
\beq
Q_{a|j}=\left(
\begin{array}{cccc}
i & 0 & 0 & 0 \\
0 & -i & 0 & 0 \\
0 & 0 & i & 0 \\
0 & 0 & 0 & -i \\
\end{array}
\right)\;.
\eeq
This implies that $\bQ_i$ and $\bP_a$ are essentially equal in this limit due to \eq{allQQ22}:
\beq
\bQ_1=i\bP_4\;\;,\;\;\bQ_2=i\bP_3\;\;,\;\;\bQ_3=i\bP_2\;\;,\;\;\bQ_4=i\bP_1\;.
\eeq
This makes our calculations much easier as we can write the gluing condition \eq{gluingF} directly on $\bP$
\beqa\la{gluing1}
\tilde\bP_4&=&-\alpha\bar \bP_2\\ \la{gluing2}
\tilde\bP_3-\bar{\alpha}\bar\bP_1&=&-\left[\beta_1+\beta_2\cosh(2\pi u)-\beta_3\sinh(2\pi u)\right]\bar \bP_4-\gamma \bar\bP_2\;.
\eeqa
To solve these equations we have to impose the asymptotic on $\bP_a$.
For simplicity we consider the $L=2$ case only, leaving general $L$ as an exercise. For $L=2$ \eq{Pasm} gives: 
\beq
\bP_1\simeq A_1 \frac{1}{u^2}\;\;,\;\;
\bP_2\simeq A_2 \frac{1}{u}\;\;,\;\;
\bP_3\simeq A_3 \;\;,\;\;
\bP_4\simeq A_4 u\;.
\eeq
Since $\bP_a$ is a function with only one branch cut which can be resolved with the help of the Zhukovsky variable $x(u)=\frac{u+\sqrt{u^2-4g^2}}{2g}$ we can use the following general ansatz\footnote{which can be interpreted as a Laurent series expansion in $x$ plane, where the functions $\bP$ are analytic in the exterior of the unit circle and the first singularity lies inside the unit circle ensuring good convergence of the series expansion.}
\beqa\la{ansatz}
\bP_1=\sum_{n=2}^\infty\frac{c_{1,n}}{x^n}\;\;,\;\;
\bP_2=\sum_{n=1}^\infty\frac{c_{2,n}}{x^n}\;\;,\;\;
\bP_3=\sum_{n=0}^\infty\frac{c_{3,n}}{x^n}\;\;,\;\;
\bP_4=\sum_{n=-1}^\infty\frac{c_{4,n}}{x^n}\;\;.
\eeqa
Note that under analytic continuation $\tilde x=1/x$.
Now we can use the condition \eq{gluing1} to deduce $\bP_4$ and
$\bP_1$. Plugging the ansatz \eq{ansatz} into \eq{gluing1} we get
\beq
\frac{c_{4,-1}}{x}+c_{4,0}+{c_{4,1}}{x}+{c_{4,2}}{x^2}+\dots=-\alpha\left(
\frac{\bar c_{2,1}}{x}+\frac{\bar c_{2,2}}{x^2}+
\frac{\bar c_{2,3}}{x^3}+\dots
\right)\;.
\eeq
We see that the l.h.s. contains infinitely many positive powers of $x$
whereas in the r.h.s. there are only negative powers, which implies that
$c_{4,n\ge 0}=0$ and $c_{2,n\ge2}=0$ and thus
\beq
\bP_2=\frac{c_{2,1}}{x}\;\;,\;\;\bP_4=-\alpha\bar c_{2,1}x\;.
\eeq
In order to deal with the second equation is a similar way we should use the identities
\beq\la{ident}
\cosh(2\pi u)=\sum_{n=-\infty}^\infty I_{2n}\left(\sqrt{\lambda}\right)x^{2n}(u)\;\;,\;\;
\sinh(2\pi u)=\sum_{n=-\infty}^\infty I_{2n+1}\left(\sqrt{\lambda}\right)x^{2n+1}(u)
\eeq
where $I_n(z)$ is the Bessel function of 2nd kind defined as
\beq
I_n(y)=\oint \frac{e^{y/2(z+1/z)}}{z^{1-n}}\frac{dz}{2\pi i}\;.
\eeq
\newpage   
\begin{exercise}
Prove identities \eq{ident}, you will have to use that $u=g(x+1/x)$
and that $g=\frac{\sqrt{\lambda}}{4\pi}$.
\end{exercise}
After that we can express both sides of \eq{gluing2} as a power series in $x$ and match the coefficients. In particular comparing the coefficients of $x^0$ and $x^{-2}$ we get
\beq
c_{3,0}=-\alpha\beta_3 c_{2,1}I_1(\sqrt{\lambda})\;\;,\;\;c_{1,2}=
\frac{\bar\alpha\bar{\beta}_3 \bar c_{2,1}}{\alpha }
I_3(\sqrt{\lambda
})\;.
\eeq
Finally, forming the combinations
\beqa
A_1 A_4&=&g c_{4,-1}c_{1,2}=-g\bar{\alpha}\bar{\beta_3}\bar c^2_{2,1} I_3(\sqrt{\lambda})\\
A_2 A_3&=&g c_{2,1}c_{3,0}=-g\alpha\beta_3 c_{2,1}^2I_1(\sqrt{\lambda})
\eeqa
which are also given in \eq{AAvals} in terms of a real quantity $e=(\Delta-L)/S$ and $S$ we conclude that
\beq
\Delta-L=S \frac{I_1(\sqrt{\lambda})+I_3(\sqrt{\lambda})}{I_1(\sqrt{\lambda})-I_3(\sqrt{\lambda})}\;.
\eeq
reproducing the result from \cite{Basso:2011rs}!
\begin{exercise}
Repeat the above calculation for arbitrary $L$. You have to obtain
\beq
\Delta-L=S\frac{\sqrt\lambda I_{L+1}(\sqrt\lambda)}{L I_{L}(\sqrt\lambda)}\;.
\eeq
In the derivation you can assume that $\gamma=\beta_1=\beta_2=0$.
We explain below why that is the case for $L=2$.
\end{exercise}
In order to fix the solution for $\bP_a$ we notice that we 
also get
\beq\la{c21}
c_{2,1}^2=\frac{iS}{g\alpha\beta_3(I_1(\sqrt{\lambda})-I_3(\sqrt{\lambda}) }\;.
\eeq
Even though the constants $\gamma,\beta_1$ and $\beta_2$ did not enter into the calculation, leading to the dimension $\Delta$, they will still appear in the solution for $\bP_a$. Here we will fix them further from the reality conditions.
Let us also show that $\gamma=\beta_2=0$. First the coefficients $x$ and $1/x$ from \eq{gluing2} give the following combination
\beq
\beta_1=-\beta_2 I_0(\sqrt{\lambda})\;\;,\;\;\gamma=-\alpha\beta_2 \frac{c_{2,1}}{\bar c_{2,1}}I_2(\sqrt{\lambda})\;.
\eeq
Since $c_{2,1}$ is already fixed \eq{c21} we obtain
\beq\la{gamma}
\frac{\gamma^2|\beta_3|^2}{|\alpha|^2I^2_2(\sqrt{\lambda})}=-\beta_2^2\bar\beta_3^2
\eeq
where the l.h.s. is real and positive. At the same time if we compare the coefficients of $x^2$ and $x^3$ in \eq{gluing2} we get
\beq\la{c3oc2}
\frac{c_{3,3}}{c_{3,2}}\frac{I_1(\sqrt{\lambda})}{I_2(\sqrt{\lambda})}=\frac{\beta_2}{\beta_3}=\frac{\bar\beta_2}{\bar\beta_3}
\eeq
where again the l.h.s. should be real due to the complex conjugation property of $\bP_a$ which allows us to complex conjugate the r.h.s.\footnote{Under the complex conjugation $\bP_a\to e^{i\phi_a}\bP_a$, for some real $\phi_a$. This implies that the ratios $c_{a,n}/c_{a,m}$ are real.}. Squaring both sides of \eq{c3oc2} and multiplying by
\eq{gamma} we obtain
\beq\la{gamma}
\frac{\gamma^2|\beta_3|^2}{|\alpha|^2I^2_2(\sqrt{\lambda})}=-\beta_2^2\bar\beta_2^2\;,
\eeq
which is only possible if $\beta_2=\gamma=0$. 

\chapter{Solving QSC at finite coupling Numerically}
\section{Description of the Method}

In this part of the notes we describe the numerical algorithm and analyze some of the numerical results. 
We illustrate the general method initially proposed in \cite{Gromov:2015wca} by considering the same states as in the previous section ${\rm tr} ZD_-^S Z$ i.e. twist-$2$ operators.
First we consider $S=2$ case -- Konishi operator.
Additionally from the beginning we impose the parity symmetry which this states have i.e. symmetry under $u\to -u$ which reflects in the parity of $\bP_a$ functions. The {\it Mathematica} code which we used for this lecture can be found as an ancillary file for the arXiv submission 1504.06640.

Below we describe the main steps and ideas for the numerical procedure.
\begin{figure}[t]
\begin{center}
\def\svgwidth{0.6\textwidth}
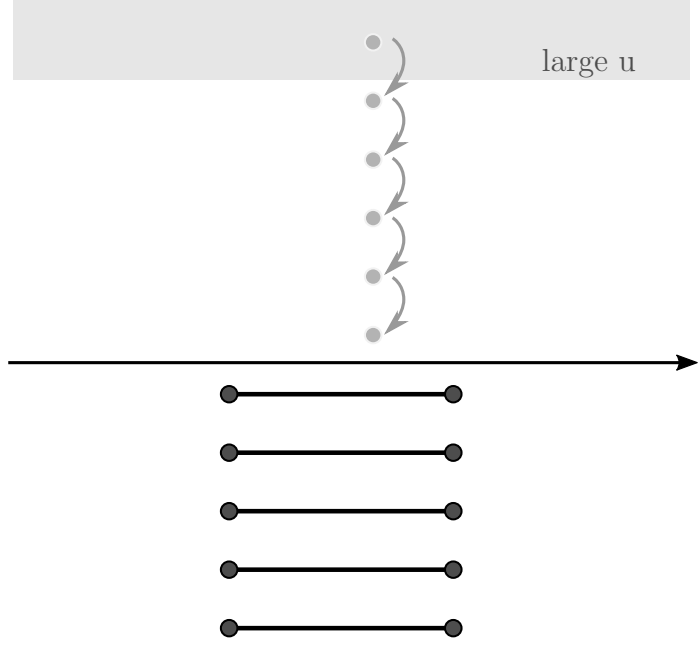
\end{center}
\caption{\la{jumps}To reconstruct $Q_{a|i}$ at the values of $u\sim 1$ we perform several jumps by $i$ using \eq{Qaijump} into the region on the upper half plane where the asymptotic expansion \eq{expansionQai} is applicable.}
\end{figure}
\begin{itemize}
\item Parameterise the system in terms of the truncated series in $x$ of ${\bf P}_a$ as follows:
\beqa\la{Px}
\bP_a=(x g)^{-\tilde M_a}{\bf p}_a\;\;,\;\;{\bf p}_a=\left(A_a+\sum_{n=1}^\infty\frac{c_{a,n}}{x^{2n}}\right)
\eeqa
where in the code we cut the sum at some finite value \mline{Pcut}. We will see that to get $6$ digits of precision with need \mline{Pcut} as small as $3$ (for relatively small $g=1/5$). This series converges very well even for $|x|=1$. Note that under the analytic continuation to the next sheet we simply replace $x\to 1/x$ so that
\beqa\la{Ptilde}
\tilde \bP_a=(x/g)^{\tilde M_a}\left(A_a+\sum_{n=1}^\infty{c_{a,n}}{x^{2n}}\right)\;.
\eeqa

\item Given ${\bf P}_a$ in terms of $c_{a,n}$ find $Q_{a|i}(u)$ as a series expansion in large $u$:
\beq\la{expansionQai}
Q_{a|i}(u)=u^{-\tilde M_a+\hat M_j}\sum_{n=0}\frac{B_{a,i,n}}{u^{2n}}\;.
\eeq
We find the coefficients $B_{a,i,n}$ by plugging the expansion \eq{expansionQai} into the finite difference equation \eq{allQQ0}. 
The term $B_{a,i,0}$ we found before in \eq{Qkjlargeu}.
Expanding it at large $u$ we get a linear system on the coefficients $B_{a,i,n}$. 
The series \eq{expansionQai} is asymptotic and works well as far as $u$ is large enough.
In our numerical implementation we keep around $12$ terms.

\item Starting from the expansion \eq{expansionQai}
at large ${\rm Im}u$ we can move down to the real axis 
using \eq{allQQ0} in the form
\beq\la{Qaijump}
Q_{a|i}(u-\tfrac{i}{2})=\left(\delta_{a}^b-\bP_a(u)\bP_c(u) \chi^{cb}\right)Q_{b|i}(u+\tfrac{i}{2})\;.
\eeq
Applying \eq{Qaijump} recursively we can decrease the imaginary part of $u$ from the asymptotic area to
reach finite values of $u$ (see Fig.\ref{jumps}). We will mostly need values of $Q_{a|i}$
at ${\rm Im}\;u=1/2$ with $-2g<{\rm Re}\;u<2g$. 
\item
Having $Q_{a|i}$ computed we reconstruct $\bQ_i$ and $\tilde\bQ_i$ from
\beq\la{QfromP}
\bQ_i(u)=Q_{a|i}(u+i/2)\chi^{ab}\bP_b(u)\;\;,\;\;
\tilde\bQ_i(u)=Q_{a|i}(u+i/2)\chi^{ab}\tilde\bP_b(u)
\eeq
where $\bP_a$ and $\tilde \bP_a$ are given in terms of $c_{a,n}$ in \eq{Px} and \eq{Ptilde}.
\item Finally, we constrain $c_{a,n}$ from the gluing conditions \eq{gluing}. We will see that it is sufficient to impose only half of them. In our numerical implementation we build a function
\beq
F(\Delta,c_{a,n},u)=\tilde \bQ_3-\alpha \bar \bQ_1
\eeq
and then adjust $\Delta$ and $c_{a,n}$ to minimize $F(\Delta,c_{a,n},u)$ at some set of probe points $u_k\in (2g,2g)$. For this we use standard numerical optimization methods.
\end{itemize}

In the next section we give more details about the {\it Mathematica} implementation of our method. 
\section{Implementation in {\it Mathematica}}
The {\it Mathematica} notebook we describe below, with slight improvements,
can be downloaded from arXiv \cite{Gromov:2015wca}.

First we make basic definitions. We define $x(u)$ in the way to ensure
that it has only one cut $[-2g,2g]$
\begin{mathematica}
X[u_] = (u + g*Sqrt[u/g - 2]*Sqrt[u/g + 2])/(2*g); 
chi = {{0, 0, 0, -1}, {0, 0, 1, 0}, {0, -1, 0, 0}, 
{1, 0, 0, 0}}; 
\end{mathematica}
\begin{exercise}
What is the branch cut structure of the naive definition\\ \mline{X[u_]=(u+Sqrt[u^2-4g^2])/(2g)} consider also the case of complex $g$.
\end{exercise}
Next we define $\hat M$ and $\tilde M$ as in \eq{relMta}. We also specialize to the Konishi operator in the $sl(2)$ sector with $J_1=S=2$. The variable \mline{d} denotes the full dimension $\Delta$; 
\begin{mathematica}
J1 = 2; J2 = 0; J3 = 0; S1 = 2; S2 = 0;
Mt = {(J1+J2-J3+2)/2,(J1-J2+J3)/2,(-J1+J2+J3+2)/2,(-J1-J2-J3)/2}
Mh = {(d-S1-S2+2)/2,(d+S1+S2)/2,(-d-S1+S2+2)/2,(-d+S1-S2)/2}
powp = -Mt;
powq = Mh - 1;
(*setting the value for the coupling*)
g = 1/5;
\end{mathematica}
The variables \mline{powp} and \mline{powq} give the powers of $\bP_a$ and $\bQ_i$.
We also set the coupling to a particular value $g=1/5$.

\paragraph*{Parameters} There are several parameters which are responsible for the precision of the result. 
\begin{mathematica}
cutP = 3;(* number of terms we keep in the expansion of P_a *)
cutQai = 12;(* number of powers in expansion of Qai at large *)
shiftQai = 20;  (* Number of jumps from asymptotic region *)
WP = 50;(* Working precision *)
PO = 12;(* Number of the sampling points on the cut to use *)
\end{mathematica}
\paragraph*{Ansatz for $\bP_a$ and parameters of the problem}
The set of ${\bf p}_a$ from \eq{Px} we define as follows
\begin{mathematica}
ps = {A[1] + I*Sum[c[1, n]/x^(2*n), {n, cutP}], 
      A[2] + I*Sum[c[2, n]/x^(2*n), {n, cutP}], 
      A[3] + Sum[c[3, n]/x^(2*n),   {n, cutP}], 
      A[4] + Sum[c[4, n - 1]/x^(2*n), {n, 2, cutP + 1}]};
\end{mathematica}
Note that we set the first sub-leading coefficient in $\bP_4$ to zero,
this is always possible to do due to the residual symmetry (see footnote \ref{rescaling}).
Whereas the coefficients $c_{a,n}$ will serve as parameters in the optimization problem, the leading coefficients $A_a$ are fixed in terms of the quantum numbers of the state via \eq{AAsl2} 
\begin{mathematica}
A[1]=-I Product[(Mt[[1]]-Mh[[j]])/If[j==1,1,Mt[[1]]-Mt[[j]]],{j,4}]
A[2]=+I Product[(Mt[[2]]-Mh[[j]])/If[j==2,1,Mt[[2]]-Mt[[j]]],{j,4}]
A[3] = 1; A[4] = 1;
\end{mathematica}
Similarly we code the leading coefficients of $\bQ_i$ and $Q_{a|i}$
\begin{mathematica}
B[1]=-I Product[(Mh[[1]]-Mt[[j]])/If[j==1,1,Mh[[1]]-Mh[[j]]],{j,4}]
B[2]=+I Product[(Mh[[2]]-Mt[[j]])/If[j==2,1,Mh[[2]]-Mh[[j]]],{j,4}]
B[3] = 1; B[4] = 1;
(* leading order coefficients in Q_ai *)
Do[B[a,i,0] = -I(A[a]B[i])/(powq[[i]]+powp[[a]]+1),{a,4},{i,4}]
\end{mathematica}
The whole Q-system, which we are partially going to reconstruct,
is thus parameterized by a set of  $c_{i,n}$ and $d$. The substitute \mline{sb} will replace these variables by their values stored in the list \mline{params}
\begin{mathematica}
prm := {d}~Join~Flatten[Table[c[i, n], {i, 4}, {n, cutP}]];
sb := Rule @@@ (Transpose[{prm, SetPrecision[params, WP]}])
\end{mathematica}
We will update the list \mline{params} at each iteration with its better approximation. As we are going to solve it with a Newton-like method, 
which is very sensitive to the starting points, one should roughly know where to look for the solution. A perturbative solution, available in some cases, could be good to start with, but sometimes even a very rough estimate of $d$ and a few first coefficients will lead to a convergent procedure. For \mline{cutP}$=3$ we need in total $1+4*3=13$ parameters. 

\paragraph*{Finding $Q_{a|i}$ at large $u$}
Having finished with defining the basics we can finally accomplish the first step in the algorithm -- find the large $u$ expansion of $Q_{a|i}$ in the form \eq{expansionQai}. First we re-expand $\bP_a$ at large $u$:
\begin{mathematica}
psu = Series[(g x/u)^powp ps/.x->X[u],{u,Infinity,cutQai+2}];
\end{mathematica}
Next we define separately the non-integer power $u^{-\tilde M_a+\hat M_j}$
and the series in inverse negative powers \eq{expansionQai}
\begin{mathematica}
qaipow = Table[u^powq[[i]]*u^powp[[a]]*u, {a, 4}, {i, 4}]; 
Bpart = Table[Sum[B[a,i,n]/u^(2*n),{n,0,cutQai/2}],{a,4},{i,4}];
\end{mathematica}
For optimization purposes we pre-expand these parts of the expansion separately with shifts $u\to u\pm i/2$
\begin{mathematica}
powP=Series[(qaipow/.u->u+I/2)/qaipow/.(u+a_)^(b_):>u^b*(1+a/u)^b
      , {u, Infinity, cutQai + 2}]; 
powM=Series[(qaipow/.u->u-I/2)/qaipow/.(u+a_)^(b_):>u^b*(1+a/u)^b
      , {u, Infinity, cutQai + 2}]; 
BpartP = Series[Bpart /. u -> u + I/2, {u, Infinity, cutQai + 2}]; 
BpartM = Series[Bpart /. u -> u - I/2, {u, Infinity, cutQai + 2}];
\end{mathematica}
Finally we code the function which computes the coefficients $B_{a,i,n}$
\begin{mathematica}
FindQlarge := Block[{},
PP=Series[KroneckerProduct[psu,chi.psu],{u,Infinity,cutQai+2}]/.sb; 
eqs=ExpandAll[Series[Normal[
(BpartP/.sb)*powP-(BpartM/.sb)*powM+(1/u)*PP.(BpartP*powP) /. sb]
,{u,Infinity,cutQai+2}]]; 
slB = Last[Solve[LogicalExpand[eqs == 0]]];
Qailarge = qaipow*Bpart/.slB/.sb]
\end{mathematica}
The function computes the expansion and store it in the variable \mline{Qailarge}.

\paragraph*{Finding $Q_{a|i},\;\bQ_i$ and $\tilde{\bQ}_i$ on the real axis}
In order to impose the gluing conditions on the Zhukovsky branch cut \eq{gluing} we will use a set of sampling points (\mline{points}), chosen so that their density increases near the ends of the interval $[-2g,2g]$ to guarantee maximal efficiency (we use Chebyshev nodes).
\begin{mathematica}
points = N[Table[-2*g*Cos[Pi*((n - 1/2)/PO)], {n, PO}], WP]; 
\end{mathematica}
Now for each of the sampling points we have to climb up to the asymptotic region using \eq{Qaijump}.
\begin{mathematica}
SolveQPP[n0_] := Block[{}, Clear[Qai, PP, PS]; 
PS[uu_] := PS[uu] = SetPrecision[Expand[(x*g)^powp*ps/.sb]
    /.x^(a_.)->X[uu]^a /. sb, WP]; 
PP[(uu_)?NumericQ] := PP[uu] = IdentityMatrix[4]+
    KroneckerProduct[PS[uu], chi.PS[uu]]; 
Qai[n0][uu_] = Qailarge /. u -> uu + I*n0 - I/2; 
Qai[n_][u_] := Qai[n][u] = SetPrecision[PP[u+I*n].Qai[n+1][u],WP]; 
Qaiplist = Table[Qai[1][p], {p, points}]]; 
\end{mathematica} 
This function creates \mline{Qaiplist} which contains values of $Q_{a|i}$
at the sampling points. This allows us to compute $\bQ_i$ using simple matrix multiplication via \eq{QfromP}
\begin{mathematica}
DoQlist := Block[{}, 
Qilist = Transpose[Table[((x*g)^powp*ps/.x->X[u]/.sb
    /.u->points[[i]]).chi .Qaiplist[[i]], {i, PO}]];
Qitlist = Transpose[Table[((x*g)^powp*ps/.x->1/X[u]/.sb
    /.u->points[[i]]).chi.Qaiplist[[i]], {i, PO}]];]; 
\end{mathematica}
Now when we have the values of $\bQ_i$ and also $\tilde \bQ_i$
we can define the function $F$, which depends on the parameters $d,c_{a,n}$ and computes the mismatch of the gluing condition
at the sampling points.
\begin{mathematica}
F[Plist_List] := (F[Plist] = Block[{}, Print[Plist]; 
params = Plist; 
FindQlarge; 
SolveQPP[shiftQai]; 
DoQlist; 
C1list = Qilist[[1]]/Conjugate[Qilist[[3]]]; 
C2list = Qitlist[[1]]/Conjugate[Qitlist[[3]]]; 
c = Mean[Join[C1list, C2list]]; 
Flatten[{Re[{C1list-c, C2list-c}/c], Im[{C1list-c, C2list-c}/c]}]]
)/; NumericQ[Total[Plist]]; 
\end{mathematica}
Finally, we have to tune the values of parameters
so that the square of the function $F$ is minimized.
\begin{mathematica}
(*setting the starting configuration*)
params0 = SetPrecision[{4.5,0,0,0,-1,0,0,1,0,0,0,0,0}, WP];
(*finding optimal parameters*)
FindMinimum[(1/2)*F[prm].F[prm], 
  Transpose[{prm, params0}], 
  Method -> {"LevenbergMarquardt", "Residual" -> F[prm]}, 
  WorkingPrecision -> 30, 
  AccuracyGoal -> 7]
\end{mathematica}
The built-in function \mline{FindMinimum} is rather slow and takes around $10$min to run. It is much better to use the implementation from the 
notebook attached to the arXiv submission \cite{Gromov:2015wca} which uses parallel computing and gives the result in about $1$ minute. It is possible to further improve the above basic code performance by roughly a factor of $10-100$, but that will also make it more cumbersome.
\begin{exercise}
Use the above code to get the dimension of the Konishi operator
at $g=1/5$. Compare your result with the high precision evaluation $\Delta=4.4188598808023509$ taken from \cite{Gromov:2015wca}.
\end{exercise}
\begin{exercise}
Use the result for $g=1/5$ as a starting point to compute $g=3/10$.
You should get $\Delta=4.826949$. Note that the convergence radius of the
perturbation theory is $g^*=1/4$\footnote{The finite convergence  radius of the perturbation theory is due to the branch-cut singularity of the spectrum at $g_*=\pm i/4$. This is the value of the coupling when branch points of the Zhukovsky cuts $2g+i n$ and $-2g+i n\pm i$ become equal.}, so this value is already outside the range accessible with   perturbation theory. 
\end{exercise}
\begin{exercise}
Check that the same code will work perfectly for non-integer values of the Lorentz spin $S$. Analytic continuation in the spin is very important for the BFKL applications \cite{Alfimov:2014bwa,Gromov:2015vua}. Try to change $S=S_1$ gradually until it reaches $S=3/2$ for $\Delta=2/10$. You should get $\Delta=3.85815$. Verify numerically \eq{gluingF} and show that $\gamma\simeq 0.0030371$ is indeed a real constant and $\beta=\beta_1+\beta_2\cosh(2\pi u)+\beta_3\sinh(2\pi u)$ for some constants $\beta_k$. 
\end{exercise}

\chapter{Applications, Further Reading and Open Questions}

In this section we attempt to cover most of the recent applications of the QSC methods and offer some open questions.

\paragraph*{QSC for ABJ(M) Theory} The QSC was also developed for ABJ(M) theory
(which is a 3D ${\mathcal N}=6$ Chern-Simons theory) in \cite{Cavaglia:2014exa,Bombardelli:2017vhk}.
An nontrivial specific feature of this theory is that the positions of the branch points are related to the `t Hooft coupling in a very nontrivial way and is called the interpolation function $h(\lambda)$. By comparing the results of localization with 
the analytic calculation of the slope function (similar to what we did in Section.\ref{slope}), it was possible to obtain an expression for the interpolation function for ABJM theory~\cite{Cavaglia:2016ide} and for a more general ABJ theory~\cite{Gromov:2014eha}. The detailed proof of these expressions is still an open question and would likely require the QSC formulation for the cusped Wilson-line in these theories, which is not known yet.

\paragraph*{QSC for Wilson Line with a Cusp}
The anomalous dimension for the Maldacena-Wilson line with a cusp was shown to be integrable in \cite{Drukker:2012de,Correa:2012hh}. In \cite{Gromov:2015dfa} the QSC construction for this observable was formulated, which allowed for the precise numerical analysis and non-perturbative analytic results. In \cite{Gromov:2016rrp} by taking an appropriate limit of the cusp anomalous dimension, the potential between heavy quark--anti-quarks was studied in detail with the help of QSC.

\paragraph*{QSC for High Order Perturbative Expansion}
The QSC method allows for a very efficient analytic perturbative expansion. A very nice and powerful method for $sl(2)$ sector was developed in~\cite{Marboe:2014gma} allowing the computation of $10$-loops analytical coefficients on a standard laptop in just $3$ hours. An alternative method, which can be applied in general situation was developed in~\cite{Gromov:2015vua}. In~\cite{Marboe:2017dmb} the project of creating a database perturbative expansion of low lying anomalous dimensions was initiated.

\paragraph*{QSC for QCD Pomeron} As we discuss in the Section~\ref{sec:cont} the QSC enables a very simple analytic continuation in the quantum numbers such as Lorentz spin $S$.
As was explained in~\cite{Kotikov:2007cy} one can approach the regime,
where ${\mathcal N}=4$ SYM becomes similar to QCD. This regime can be also studied with the QSC~\cite{Alfimov:2014bwa}. In particular the most complicated highest transcedentality parts of the planar QCD result
at $3$ loops was obtained for the first time 
in~\cite{Gromov:2015vua}, by using the QSC. It was later confirmed by an independent calculation in~\cite{Caron-Huot:2016tzz}.

\paragraph*{QSC for Deformations of ${\mathcal N}=4$ SYM}
The ${\mathcal N}=4$ admits numerous deformations. Some of them are
analogous to the twists we discussed in Section~\ref{sec:twist} and can be easily introduced into the QSC formalism simply by modifying the asymptotic of the Q-functions. For some examples see~\cite{Gromov:2015dfa,Kazakov:2015efa}. Another deformation is
called $\eta$-deformation~\cite{Arutyunov:2013ega}, which most likely can be described by the QSC as well, by replacing a simple cut in $\bP_a$ function with a periodised set of cuts\footnote{this case was considered very recently in~\cite{Klabbers:2017vtw}.}.

\paragraph*{QSC for Fishnet Graphs}
In the limit when one of the twist parameters becomes large and the 't Hooft coupling simultaneously scales to zero one gets a significant simplification
in the perturbation theory, which gets dominated by the ``fishnet" scalar graphs. First this limit was considered for the cusp anomalous dimension in~\cite{Correa:2012hh,Erickson:1999qv} and it was possible to reproduce the result analytically from the QSC. 
A more systematic study of the ``fishnet" limit of ${\mathcal N}=4$
was initiated by~\cite{Gurdogan:2015csr} where it was demonstrated that
many more observables can be studied by considering a special type of diagram. In~\cite{Gromov:2017cja} it was shown how the QSC methods can be used to evaluate these type of Feynman graphs.

\paragraph*{Open Questions} Even though a number of longstanding problems were resolved with the help of the QSC there are still a number of open questions which could potentially be solved using the QSC. Some of them are likely to be solved soon, others may never be solved. Below we give an incomplete list of such problems, focusing on those more likely to be solved before the next ice age. 

It would be very useful to be able to extract the strong coupling expansion of the spectrum analytically from the QSC. Some first steps were done in~\cite{Hegedus:2016eop}. 

The structure of the QSC is very constraining and at the moment we only
know two QSCs for SYM and ABJ(M). It would be useful to make a complete classification of the QSCs starting from the symmetry group.
 This way one should find the QSC for $AdS_3/CFT_2$ and also possibly for a mysterious $6D$ theory -- a mother theory of 6D integrable fishnet graphs. Similarly, different asymptotic and gluing conditions  represent different observables in ${\mathcal N}=4$ SYM, it would be
 useful to have complete classification of such asymptotics and gluing conditions.
For even more mathematically oriented readers there is the question of
proving existence/countability of the solution of the QSC.

A big open conceptual question is how 
to derive the QSC from the gauge theory perspective, without a reference to AdS/CFT correspondence as this would allow
us to prove to some extent AdS/CFT by taking the classical limit of the QSC, deriving the Green-Schwartz classical spectral curve.

Some of the problems which are within immediate reach include:
studying oderon dimension in the way similar to BFKL pomeron~\cite{Gromov:2015vua} (see for settings~\cite{Brower:2014wha});
constructing the QSC for the recently proposed integrability framework for the 
Hagedorn phase transition~\cite{Harmark:2017yrv} which would enable analytic weak coupling expansion and numerical analysis for this observable;
integrable boundary problems with   
non-diagonal twist, like recently considered in~\cite{Guica:2017mtd}
could be most likely treated in the way similar  to~\cite{Gromov:2016rrp}, this problem seems to be also related to the 
problem of finding the spectrum of tachyons~\cite{Bajnok:2013wsa}, which is another problem where the QSC reformulation could help to advance further.

A more complicated but very important problem is to extend the QSC formalism to the problem of computing n-point correlation function.
Existing beautiful integrability-based hexagon formalism~\cite{Basso:2015zoa} should give important hints on re-summing wrapping corrections. This problem seems to be linked to the problem of finding separated variables in the AdS/CFT for some first steps at weak coupling see~\cite{Gromov:2016itr}. The 
one-point function~\cite{Buhl-Mortensen:2015gfd} could be the perfect framework for developing the new QSC-based formalism for the correlators.
See also~\cite{Buhl-Mortensen:2017ind} for more exotic observables which could be also potentially governed by integrability. 

Finally, the other main open questions are whether we could also use integrability to get non-planar corrections and also get closer to the real world QCD.
\\{ }\\
If you have questions, please feel free to email to \url{nikgromov@gmail.com}. You are also welcome to email any answers to the above questions to \url{nickgromov@mail.ru}!

\end{document}